\documentclass[structabstract]{aa}  
\usepackage{graphicx}
\usepackage{txfonts}
\begin{document}
   \title{A non-hydrodynamical model for acceleration of line-driven winds in Active
Galactic Nuclei}

%   \subtitle{I. Overviewing the $\kappa$-mechanism}

   \author{G. Risaliti \inst{1,2} 
          \and
          M. Elvis\inst{2}
          }

   \institute{
             INAF-Osservatorio Astrofisico di Arcetri, Largo E. Fermi 5, 50125 Firenze, Italy             
              \email{risaliti@arcetri.astro.it}
         \and
Harvard-Smithsonian Center for Astrophysics, 60 Garden Street, Cambridge, MA 02138  }
   \date{xxx }

% \abstract{}{}{}{}{} 
% 5 {} token are mandatory
 
  \abstract
{Radiation driven winds are the likely origin of AGN outflows, and are believed to be a fundamental 
component of the inner structure of AGNs. Several hydrodynamical models have been developed, showing that these winds can be effectively launched from AGN accretion discs.}
{Here we want to study the acceleration 
phase of line-driven winds in AGNs, 
in order to examine the physical conditions for the existence
of such winds for a wide variety of initial conditions.
 }
{We built a simple and fast non-hydrodynamic model QWIND, where we 
assume that a wind is launched from the accretion disc at supersonic 
velocities of the order of a few 10$^2$~km/s and we concentrate 
on the subsequent supersonic phase, when the wind is  
accelerated to final velocities up to 10$^4$~km/s.  }
{We 
show that, with a set of initial parameters in agreement with 
observations in AGNs, this model can produce a wind with terminal 
velocities of the order of $10^4$~km/s.
% , similar to BALs, and 
%with a structure quite similar to that found in more complex hy- 
%drodynamical models. 
There are three zones in the 
wind, only the middle one of which can launch a wind: in the 
inner zone the wind is too ionized and so experiences only the 
Compton radiation force which is not effective in accelerating 
gas. This inner ‘failed wind’ however plays an important role in 
shielding the next zone, lowering the ionization parameter there. 
In the middle zone the lower ionization of the gas leads to a 
much larger radiation force and the gas achieves escape velocity 
This middle zone is quite thin (about 100 gravitational radii). 
The outer, third, zone is shielded from the UV radiation by the 
central wind zone and so does not achieve a high enough 
acceleration to reach escape velocity. 
We also describe a simple analytic approximation of our 
model, based on neglecting the effects of gravity during the
acceleration phase. This analytic approach is in agreement with 
the results of the numerical code, and is a powerful way to check 
whether a radiation driven wind can be accelerated with a given 
set of initial parameters.}
   {Our analytical analysis and the fast
QWIND model are in agreement with more complex hydrodynamical models, 
and allow an exploration of the dependence of 
the wind properties for a wide set of initial parameters: black hole mass, 
Eddington ratio, initial density profile, X-ray to UV ratio.
}


   \maketitle
%
%________________________________________________________________

\section{Introduction}

Outflowing winds are now believed to be common, and quite possibly ubiquitous, 
in the inner parts of
Active Galactic Nuclei (AGNs) and quasars. The most striking evidence
comes from the 10\% of Broad Absorption Line (BAL) quasars which show
broad absorption lines blueshifts spanning 10-20 thousands
km~s$^{-1}$.  Less spectacular, but more common, evidence for outflows
comes from the 50\% of AGN with Narrow Absorption Lines (NALs) both UV and
X-ray (Reynolds 1997, George et al. 1998, Crenshaw et al. 1999, 
Krongold et al. 2003, Vestergaard 2003, Piconcelli et al. 2005, Ganguly \& Brotherton~2007).

Several attempts have been made to explain the origin of such
winds. Two main scenarios have been suggested: magnetically driven
winds (Blandford \& Payne 1982, Konigl \& Payne 1994, Konigl \& Kartje~1994,
Everett \& Murray~2007), or radiation
driven winds.  One of the most promising explanations is through
radiation line-driven winds arising from the accretion disc. The
physics of radiation line-driven winds based on the work of Sobolev
(1960) was developed by Castor, Abbott \& Klein (1975, hereafter CAK),
and Abbott (1982, 1986) for winds from hot stars. CAK showed that
resonance line absorption in an accelerating flow could be hundreds of
times more effective than pure electron scattering. More recently the same
theory has been applied to accretion discs in compact binaries (
Proga,
Drew \& Stone 1998) and to AGN discs (Murray et al.~1995, 
Proga et al. 2000, hereafter P00, 
Proga 2003). 
These authors use the powerful hydrodynamical code ZEUS2D (Stone \&
Norman 1992) to solve the wind equations, assuming a Shakura-Sunyaev
(1972, hereafter SS) $\alpha$-disc, and plausible initial conditions.
The main results of these works are (1) the demonstration that a wind
can be launched and accelerated up to velocities of the order of
10$^4$~km~s$^{-1}$, and (2) the determination of the wind geometry and
physical state for the given starting conditions.
%A limitation of the detailed approach however is that the very
%complexity of the modelling both makes it hard to discern the key
%physical processes, and also that run times for the code become so long
%that the parameter space cannot be thoroughly explored. The
%possibilities for gaining a physical understanding for the system are
%thereby reduced.
These models have been tested 
with several different choices of the 
initial conditions (Proga \& Kallman~2004), showing that a wind can arise
for a wide range of black hole mass and accretion rates. 

Further recent improvements in this field are
the modeling of Schurch \& Done~(2007), where the
interaction between the X-ray radiation and the outflowing wind is
throughly analyzed, and the analysis of possible effects of radiation
winds at larger distances from the accretion disc, such as on a parsec-scale
X-ray heated torus (Dorodnitsyn et al.~2008), and on large-scale AGN outflows
(Kurosawa \& Proga~2009).

%
%Furthermore, a purely SS disc is assumed, and only radiation pressure
%is taken into account in the initial launching phase of the wind, in
%the region just above the accretion disc atmosphere.  The current
%uncertainty regarding the physical condition in the disc atmosphere
%makes these assumptions reasonable, but not unique.  A different disc
%structure, the presence of magnetic fields and the effect of the
%high-energy radiation can significantly alter this scheme and produce
%significantly different initial conditions.

One key aspect of the physics of radiation driven winds, clearly emerging 
from the models mentioned above, is that 
regardless of the details of the initial launching phase, it
is impossible to avoid the strong, probably dominant, effect of
radiation pressure in the subsequent acceleration phase, where the
wind gains more than 99\% of its kinetic energy. This is easily
estimated from the comparison between the amount of momentum absorbed
by the gas and its final momentum. For example, Hamann (1998)
estimates that at least $\sim25$\% of the UV radiation emitted by the
Broad Absorption Line Quasar PG~1254+047 is absorbed by a gas with
column density of the order of 10$^{23}$ cm$^{-2}$. It is sufficient
that the luminosity of this source is of the order of 10\% of the
Eddington luminosity to conclude that the momentum in the outflowing
wind is of the order of that absorbed in the UV wavelength range.

The distinction between the {\em launching} and {\em acceleration}
phases of an accretion disc wind, which we made above, allows the
modelling of the problem to be separated into these two parts.  
{\bf The {\em launching} phase has been investigated in several numerical
simulations, both in hydrodynamical (Ohsuga et al.~2005) and 
magneto-hydrodynamical (Hawley \& Krolik~2006) regimes. Recently,
the creation of outflows from accretion discs has been investigated through
MHD simulations including the effects of radiation pressures (Ohsuga et al.~2009).
}

The
main aim of the work presented here is to explore a wide set of initial
conditions for the {\em acceleration} phase of a wind in AGN  using a deliberately
simplified approach in the hope
that this can produce an intuitive understanding of quasar winds and
so guide future detailed simulations. We will make use both of
analytic approximations and of a simplified numerical code for quasar
winds (which we named {\em QWIND}). {\em QWIND} treats in detail the
radiation force mechanism, but not the internal gas pressure in the
wind. As the wind velocity in the acceleration phase is always many
times the thermal velocity of the wind gas, this is a reasonable
approach.

Our study is motivated by the observations of fast outflows in Broad
absorption Line (BAL) quasars (Weymann 1997), by photometric evidence
for a highly flattened structure in BAL winds (Ogle et al.~1999) 
and by results suggesting
an axially symmetric, but not spherical, spatial distribution of the
Broad Emission Line (BEL) gas (Wills \& Browne 1986, Brotherton 1996,
Maiolino et al. 2001, Rokaki et al. 2003).  
%The basic idea behind our model is that the BEL gas could be embedded in
%a warmer outflowing gas. This warm outflow would be present in all quasars, but
%is observable only along particular lines of sight. 
Elvis (2000) has proposed a specific quasar unification model based on
this kind of structure.  This model predicts that a geometrically
thin, funnel-shaped outflow is present in all AGNs and quasars, which
allows the observational properties of the different classes of
sources to be explained through orientation effects. This model
provided our initial motivation, as we suspected that a radiation
driven wind might naturally create a thin, funnel-shaped structure.

We therefore developed the {\em QWIND} model in order to easily and quickly 
explore 
the acceleration of a radiation-driven wind 
for the huge range of physically possible initial conditions, in terms of black hole mass,
accretion rate, relative strength of the X-ray radiation, initial
 density and
temperature of the outflowing gas. Our approach does not add more
physical insights for the single wind solutions,
than the already available hydrodynamical models mentioned above, but
allows a much wider exploration of the initial parameters space.

The purpose of this paper is to present the model code, {\em QWIND}, and to demonstrate that {\em QWIND} 
produces results in agreement with both an analytic treatment and with the more
complex and detailed approach of P00.
The structure of this paper is the following: in Section 2 we briefly
review previous results on the topic of radiation-driven winds.  In
Section 3 we present the {\em QWIND} code and we show examples of its
possible applications.
In Section 4 we
present an analytical treatment of the wind equations, which provides
interesting constraints on the existence of such winds, as a function of
the initial parameters. In Section~5 we briefly compare our results
with those obtained by more complex hydrodynamical codes. 
Finally, in Section~6 we present our
conclusions and outline future work.

%%%%%%%%%%%%%%%%%%%%%%%%%%%%%%%%%%%%%%%%%%%%%%%%%%%%%%%%%%%%%%%%
\section{Summary of previous results}

The radiation force on a moderately ionized 
gas due to incident ultraviolet radiation
is primarily via line absorption, rather than continuum electron scattering (CAK). 
In a constant velocity gas the 1000 times larger cross-sections
of resonant line transitions over continuum Thomson scattering has little
effect because the narrow wavelength ranges spanned by each transition
contain relatively little UV continuum.  However, a real wind, driven
by a central radiation source will be accelerated, and so the wavelength for
each absorption line will be Doppler shifted by the relative velocity
of the gas with respect to the source.  If the radial velocity gradient
shifts the absorption wavelength sufficiently for each gas
element, then fresh UV continuum is absorbed, leading to continued acceleration.
As the process continues the fraction of the UV continuum that is absorbed is
greatly increased.  The absorption cross-section then becomes a {\it
local} function of the physical conditions of the wind (the ``Sobolev
approximation'', Sobolev 1960).  CAK introduced a numerical ``force
multiplier'', M(t), which represents the enhancement of the radiation
force due to line absorption with respect to pure electron scattering.
$M(t)$ depends on only one local quantity, the ``effective optical
depth'', $t$, defined as:
\begin{equation}
t=\sigma_T n v_{th} |\frac{dv_r}{dr}|^{-1}
\label{eqmt1}
\end{equation}

where $\sigma_T$ is the Thomson cross section, $n$ is the gas
density $v_{th}$ is the thermal velocity in the gas, and
$v_r$ is the radial component of the velocity.

Taking into account the thousands of, mostly weak, UV absorption lines for each ion,
CAK estimated the following analytic power-law approximation for M(t):
\begin{equation}
M(t)=K\times t^{-\alpha}
\label{eqmt2}
\end{equation}

where $\alpha$ is a positive number (with a typical value
$\alpha=0.6$) and $K$ is of the order of 0.1-0.4.

However, as a gas becomes more ionized, either collisionally due to an
increased temperature (e.g. through photoionization due to the intense
UV source), the number of transitions available for absorption is
reduced quite dramatically. In launching AGN winds this is a well
known problem, and {\em ad hoc} schemes to shield the gas have been
proposed (e.g. ``hitchhiking gas'', Murray \& Chiang 1997)

%%%%%%%%%%%%%%%%
\begin{figure}
\includegraphics[width=8.5cm]{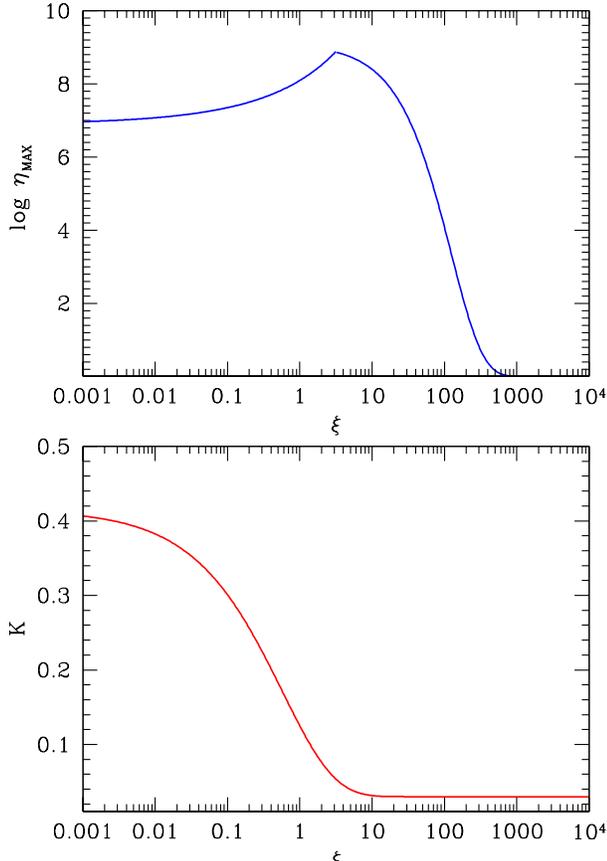} 
\caption{
Analytic approximation for
the parameters $K$ (solid line) and $\eta_{MAX}$ (dashed line), as a
function of $\xi$ (Stevens \& Kallman, 1986)\label{stevens1}}.
\end{figure}
%%%%%%%%%%%%%%%%

\begin{figure}
\includegraphics[width=8.5cm]{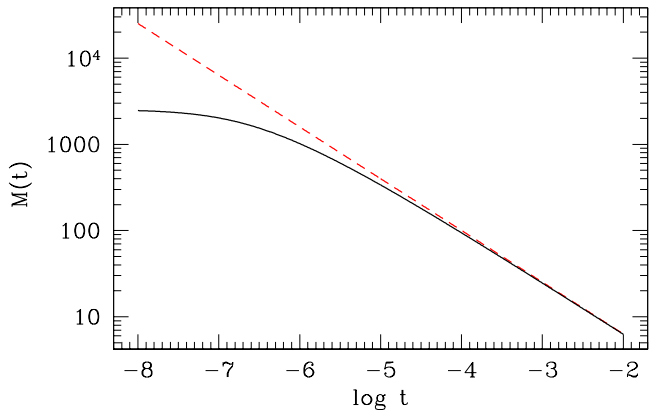} \caption{Force multiplier $M(t)$ vs optical
depth $t$ for a neutral gas in the Castor, Abbott \& Klein (1975)
formulation (straight dashed line), and the Abbott (1982) and Stevens
\& Kallman (1986) formulation that allows for ionization state and is
more accurate for low effective optical depth
($t<10^{-5}$).\label{stevens2}}
\end{figure}

A useful, second-order approximation, which takes into account the
ionization state of the gas and solves the divergence of the previous
equation for small $t$, is the following (Abbott 1986):
\begin{equation}
M(t)=K\times t^{-\alpha} \times [\frac{(1+\tau_{MAX})^{1-\alpha}-1}{\tau_{MAX}^{1-\alpha}}]
\label{eqmt3}
\end{equation}

with $\tau_{MAX}=t\times \eta_{MAX}$, where $\eta_{MAX}$ is a coefficient
that is strongly dependent on the ionization factor, $\xi$. This
relation was studied by Stevens \& Kallman (1986) who produced an
analytic approximation that is correct within 10\% over the whole
range of interest for $\xi$.  We plot this relation in
Fig.~\ref{stevens1}, together with the dependence on $\xi$ of the
constant $K$.  In a wind, $\xi$ can be as high as $10^3-10^4$ in the
inner surface, but then rapidly drops to small values ($\xi<1$)
because of the lowered X-ray flux. As a consequence, in the range of
interest for a wind both $\eta_{MAX}$ and $K$ are constant to within a
factor $\sim 2$. In Fig.~\ref{stevens2} we plot Eq. 2 and 3 for a
neutral gas. We note that the correction factor is significant only if
$t < 10^{-5}$.

%%%%%%%%%%%%%%%%
%%%%%%%%%%%%%%%%

Using these results, P00 developed a code which solves the
hydrodynamical equations for a wind arising from an accretion disc,
pushed by a central radiation force described by the above equations.
In this work, a SS disc is assumed, and the ionization state is
determined by the incident X-ray radiation from the central source.
The main conclusion of P00 (in agreement with the previous work of
Murray et al. 1995) is that a wind can easily arise from the accretion
disc of AGN, under some initial conditions (SS disc, with a choice of
free parameters such as the Eddington ratio and the X/UV ratio, which
will be discussed in the next Sections).  The dependence of the wind
properties on the black hole mass and accretion efficiency is
particularly interesting for they can now be tested against
observations, as we discuss below.  The black hole mass, $M_{BH}$
and the accretion efficiency relative to Eddington, $\epsilon_{EDD}$,
are the parameters determining the luminosity of AGN.
%
%A dependence of the wind properties on luminosity is suggested by the
%fact that broad absorption lines, blueshifted by several 10$^4$ km
%s$^{-1}$ are observed in BAL QSOs (representing $\sim$~10-15\% of the
%entire quasar population) but not in the lower luminosity Seyfert
%galaxies. 
%
%An observational property of QSOs that could be related to the wind is
%the Baldwin effect (Baldwin 1977, Osmer \& Shields 1999).  This still
%unexplained anti-correlation between luminosity and equivalent width
%of UV emission lines (primarily Ly$\alpha$ and CIV) could be due to the
%dependence of wind geometry on luminosity, if the outflowing wind contains the
%broad emission line clouds (Elvis 2000).

%For these reasons, it is particularly important to investigate the
%dependence of a radiation driven wind  on the black hole mass,
%M$_\bullet$ and the Eddington ratio, $\epsilon_{EDD}$. 

%%%%%%%%%%%%%%%%%%%%%%%%%%%%%%%%%%%%%%%%%%%%%%%%%%%%%%%%%%%%%%%%%%%%
\section{Radiation-driven acceleration of a wind} 

Several physical effects can be considered concerning  the equations of the radiation force,
to understand whether this external force is capable of
accelerating the wind up to velocities of $\sim~10^4$ km~s$^{-1}$ or
higher, as observed in BAL quasars.

The results summarized above (Section~2) show that a radiation driven wind can
arise from accretion discs in AGNs. However, the breadth of initial
conditions that produce such a wind are at present unknown.  First,
while the SS disc is a plausible solution, 
%the $\alpha$
%parametrization of viscosity is arbitrary and without a physical
%basis, and 
many other disc structures are possible.
Secondly and most importantly, even in the SS disc
paradigm several initial parameters are unknown: the black hole mass,
the 
Eddington ratio, and the viscosity parameter $\alpha$.  Finally,
since the physical reason for the initial launching of a wind is not
clear, the radius at which the wind is launched is
unknown. However, recent studies of warm absorber winds in local AGNs suggest
that winds on scales as small as a few thousand 
gravitational radii from the central black holes are possible
(Krongold et al.~2006). 
%Instabilities in the disc structure are a likely cause. In a
%SS disc one instability point lies at the boundary between radiation
%pressure- and gas pressure-dominated regions (Lightman \& Eardley
%1974).  Another instability point occurs at the hydrogen ionization
%radius (Mineshige \& Shields 1990, Siemiginowska, Czerny \& Kostyunin 1996).

An important aspect of the solutions found in the previous works of
Murray et al. (1995) and Proga et al.~(2000) is that the sonic and
critical points of the wind are reached at a small height $H$ above
the disc with respect to the distance from the central source $R$
($H/R<0.1$). Also, the velocitiies at the sonic and critical points are of
the order of a few 10$^2$~km~s$^{-1}$, which is only a few percent of the final
BAL velocities, of the order of 10$^4$~km~s$^{-1}$.

As briefly discussed in the introduction, radiation pressure is
expected to be relevant (and probably dominant) in the wind
acceleration, regardless of the details of the launching mechanism.
However, following the theory summarized in Section~2, in order for
the absorption line mechanism to be effective, a balance between the
intensity of the X-ray radiation (which determines the ionization
parameter) and the intensity of the UV radiation (which provides the
momentum needed by the gas to accelerate to high velocities) is
required (Murray \& Cheng~1995).

In this paper we discuss the general conditions for the acceleration of a
line-driven wind after the launching phase.  In particular we
investigate whether the ``thin wind'' geometry proposed by Elvis (2000), which
is successful in explaining a wide set of observational properties,
can arise naturally. The main unexpected features of the Elvis~(2000) geometry are
that (1) the wind is thin ($\Delta$R/R$<$1) and that (2) the wind is
initially quasi-vertical, making a hollow cylinder of height
$h/r\sim1$, before becoming a radial, biconical flow.\\

\subsection{The  {\em QWIND} code}

We developed the {\em QWIND} code with the following
assumptions:\\
\begin{enumerate}
\item The X-ray source is point-like, isotropic and located at the
center of the disc.
\item The UV/optical source is the accretion disc, emitting according
to the SS model. 
{\bf In our model the wind inner radius can be as small as 100~R$_S$, so 
the dimensions of the emitting region of the disc are non-negligible,
and the disc cannot be considered a point source. 
Therefore, the dependence of the disc emissivity with radius, and the
actual disc extension are taken into account.
%The emission profile with disc radius turns out not
%to be important for the final result, since the region of the disc
%emitting most of the ultraviolet force-driving radiation is small ($\sim50~R_G$)
%compared with the wind inner radius $R_{IN}$
%In contrast, 
We also take into account the
anisotropy of the disc emission, which
has important consequences for the
wind properties, the radiation being zero on the disc plane, and
increasing with the cosine of the inclination angle of the disc with
respect to the wind. Limb darkening (Fukue \& Akizuki~2007), which would accentuate this
effect, has been neglected.}
\item The gas arises from the disc at all radii between an inner radius 
$R_{IN}$ and an outer radius $R_{OUT}$, at a  
height $H$ above the disc, with a given density $n$, temperature
$T$, and initial vertical velocity $v_0$. 
The radial velocity of this gas is zero, while the angular velocity in
the disc plane is Keplerian.
\item The wind is subject only to the central gravitational force
$F_G$ due to the black hole of mass $M_{BH}$ and radiation force
$F_r$.  Internal gas pressure is neglected, as is the gravity of the
disc. 
%However, it is implicitly taken into account by requiring
%that in the vertical phase of the wind the velocity is constant until
%the radiation force becomes stronger than gravity (see discussion in
%Sect. 3). 
\item The radiation force in each element of the gas is computed using
the CAK and Abbott~(1986) equations described in Section 2 (Eqs. 2 and 3). The ionization
parameter, $\xi$, is determined assuming a cross section
$\sigma_X=100\sigma_T$ if $\xi<10^5$, and $\sigma_X=\sigma_T$ if
$\xi>10^5$. This simple approximation reflects the fact that when the
ionization factor is too high, there are no bound electrons for
photoelectric absorption to be effective. Below this critical value,
photoelectric absorption on the electrons in the inner shells of
metals is much more effective than Thomson scattering in removing
X-ray photons.
Using this approximation, if $\xi>10^5$, X-rays penetrate deep
into the wind, keeping $\xi$ high, until the Thomson optical depth is higher
than 1 ($N_H>1.5\times10^{24}$~cm$^{-2}$) 
or the $R^2$ factor makes $\xi$ decrease below 10$^5$. Until
this point, the radiation force is negligible because $M(t)$ is small.
Deeper inside the wind, where $\tau_X>1$, the UV radiation is also
fully absorbed, and again $\sigma_X=\sigma_T$, and therefore the
radiation force will be negligible in this case too.
If instead $\xi<10^5$ the X-rays are absorbed in a thin layer
\footnote{ The problem can be complicated by the effects of secondary
ionizations, which make $\xi$ decrease more slowly than exponentially
(Maloney et al. 1996). However, this effect is not important when
$\xi$ is high, and the external electrons (the targets for secondary
ionizations) are already free.}
($N_H\sim10^{22}$ cm$^{-2}$) and the ionization parameter drops as
$\xi\propto e^{-100\sigma_T N_H}$, rapidly reaching the values 
%$\xi < 100$ 
at which the force multiplier becomes significantly higher than 1.
\end{enumerate}

{\em QWIND} follows a gas element of radial thickness $\Delta$R ($\Delta$ R$<<$R)
from the accretion disc to radii
greater than $10^3~R_S$, starting from a set of initial radii
$R_i$.\footnote{Note that throughout the paper $R_i$ and $n-i$ refer to the initial radius and
density, respectively, of the i-th stream line,
while $R_{IN}$ and $n_{IN}$ are the initial radius and density of the first stream line,  
i.e. $R_{IN}$=$R_i$(i=1), $n_{IN}$=$n_i$(i=1)}
{\bf This approach, consisting of following the trajectories 
of a fluid element, is similar to that adopted in several previous
works analyzing disc outflows (Icke et al.~1980, Tajima \& Fukue~1996,
Watarai \& Fukue~1999).} 

{\bf The acceleration of a gas element belonging to the i-th 
stream line, at coordinates $R$, $\phi$, $z$, is simply given by:
$a=-GM_{BH}R^{-2}+F_{RAD}/m(R)$
where $m(R)$ is the mass of the gas element, calculated  from the
mass conservation equation (therefore depending on the radius and
velocity of the gas element, and on the initial density profile).
The radiation force $F_{RAD}$ is estimated integrating the contribution
of each disc element:
\begin{equation}
F_{RAD}=\int_{DISC}
f(\phi)e^{-\tau}\frac{l}{4\pi R^2c}\sigma_T(M(t)+1)
\end{equation}
where f($\phi$) is the geometrical projection factor of each disc element
as seen from the gas element; $l$ is the luminosity of the disc
element; M(t) is the force multiplier,
estimated as described above, and $\tau$ is the optical depth,
estimated adding the contributions of the stream lines between the gas element
and the disc: $\tau$=$\Sigma_i$$\sigma_i\times
$$\Delta$R$_i$$\times$n($R_i$)/f($\phi$), where n($R_i$) is
the density of the gas element in the i-th stream.  
The luminosity $l$ of each disc element depends on the radial distance from the center
of the disc, as estimated in a SS disc. 

Our choice to integrate over the disc structure is motivated by the 
non-negligible dimensions of the disc with respect to the inner wind
streams: in our computation the disc emission is calculated up to
an outer radius of 400~R$_S$ (even if its contribution to the total
luminosity is negligible at R$>$100~R$_S$), while the inner wind radius
is 100~R$_S$ (see next Section for further details on the computation
domain). 

The computation starts from the inner stream line (free from absorption), and then
proceeds to the outer ones. At each step, the calculations are
done in the following order:\\
- We estimate the absorption due to the inner stream lines.
For the UV component, this is done simply
adding the column densities of the inner stream lines, and multiplying
by the Thomson cross section. For the X-ray component, the computation
is more complex: for each stream line crossed by the light ray from the
disc element, we estimate the ionization factor and then, depending on
its value, the X-ray cross section, following the prescription described
above.  \\
- Using the absorption correction estimated in the previous point, and integrating
over each disc element, we
compute the flux on our wind element. \\
- We estimate the density of the gas element from the values of density, mass and velocity
of the previous computation step, and requiring mass conservation.\\
- We estimate the force multiplier M(t), following the prescription 
described in the 5th point of the code properties.  \\
- We obtain the acceleration on the gas element, and we compute
the new spatial coordinates and velocity.

As a consistency check of the numerical stability of QWIND, we
performed several runs with no radiation force contribution, and we 
checked that the evolution of the stream lines are those expected
in a Keplerian problem. In particular, we checked that the conservation 
of  angular momentum holds, and that the results are stable against
changes in the number and dimensions of the gas elements used in the computation.  

}
Neglecting the internal gas pressure makes the stream lines of
gas in {\em QWIND} intersect unphysically. When two stream lines
intersect, this means that the gas in the internal line is pushed
outwards more than the gas on the external line. In reality this
obviously cannot happen, and is prevented by the horizontal component
of the gas pressure. However, the two stream lines do delineate a limiting cone
that a fully modelled wind must keep within.  The gas pressure force
between the two ``lines'' will make the gas move along a line whose
inclination angle with respect to the disc axis is intermediate
between the two lines computed by {\em QWIND}.  This approximation
therefore makes the results of {\em QWIND} usable only in this
bounding sense, but should not change the general findings that we are
interested in, i.e. whether the wind can be accelerated by radiation
pressure, the approximate angles to the disc, and covering factor of
the central source, since these properties strongly affect the
observables.  Another important effect neglected in our treatment is
the change in momentum at the bending of the wind, which is expected
to produce shocks and, possibly, soft X-ray emission, although this is
unlikely to dominate the X-ray luminosity of the AGN. The
case for the outflowing gas as a source of the soft X-ray excess
observed in many quasars has been discussed by Pounds et al. (2003).

The number of stream lines used in the runs discussed in this paper is
kept to just 20.  We also made several runs with a larger number of
lines (50), to check for discretization effects and found that the
properties of the solutions do not change significantly.

The initial density of the wind at a radius R is one of the main free 
parameters of our model. In general, we parametrize the density as 
$n_i(R_i)=n_{IN}\times(R_i/R_{IN})^{-\beta}$. Two particularly interesting cases
are $\beta=0$ (constant density at all radii) and $\beta=15/8$, the
density scaling in the SS disc. In this paper we adopt $\beta=0$ in our examples. 
Different cases (and, in particular, the SS disc-like profile, will be investigated in 
a forthcoming paper.

The initial vertical velocity  $v_0$ is a free
parameter of the code and is of the order of a few 100~km~s$^{-1}$.
%In order to have an outflowing wind we
%need it to be a few thousand km~s$^{-1}$ at radii of
%300-500~$R_S$. 
Note that this velocity is much lower than the escape velocity,
therefore most of the kinetic energy needed to make a wind must be
provided by the external radiation force.

%Once a $v_0$ is assumed, the initial velocity profile along
%the disc is determined assuming that this is related to the luminosity
%emitted at each single radius.  In particular, we assumed that a fixed
%fraction of the energy emitted by each ring of the disc is converted
%into kinetic energy of the gas above that ring\footnote{This is an
%approximate treatment, since we neglect the force needed to support
%the gas against the vertical component of the gravity. We discuss this
%point in Appendix A.}. With this assumption, we find that the initial
%velocity profile, $v_0(R)$, is related to the initial density
%profile. We have:
%\begin{equation}
%v_0(R)=v_0(R_{IN})\times (\frac{R}{R_{IN}})^{-1+\frac{\beta}{3}}
%\end{equation}
%where $\beta$ is the exponent in the density profile equation
%(in our case $\beta=15/8$, see Section 3.1).

With these assumptions, the total mass outflowing from the disc is
easily derived, integrating the contribution from each disc ring.
%\begin{equation}
%\dot{M}_D=\frac{6\pi}{2\beta-3}\rho_0v_0R_0^2[1-(\frac{R_{OUT}}{R_{IN}})^{1-\frac{2}{3}\beta}]
%\end{equation}
%Assuming the standard set of parameters used above, and
%$R_{OUT}=5R_{IN}$ we obtain
%$\dot{M}_D\sim10^{-2}M_7^2$~M$_\odot$~yr$^{-1}$.  
The fraction of this mass that falls back onto the disc versus the
fraction which escapes through a wind, depends on the subsequent
acceleration phase, and is studied by our model.

The initial parameters $v_0$ and $n_i(R_i)$ are in reality a
result of the previous launching phase. For example, in a purely
radiation-driven wind, the mass loss rate is uniquely determined by
the regularity and stability conditions at the 'critical point' (CAK,
Lamers \& Cassinelli 1999, Murray et al.~1995).  Since we are not
studying this launching phase here, a fundamental requirement is that
our final results do not depend critically on the exact values of
these parameters. Since the density profile does affect the
acceleration phase of the wind (because it determines, together with
the X-ray flux, the ionization parameter), it is particularly
important that the final results of the acceleration phase are
independent on the exact value of the initial velocity, provided that
it is small compared with the final velocity, and large enough to be
supersonic (this, for temperatures
of a few 10$^6$~K implies $v_0$ of the order of a few
10$^2$~km~s$^{-1}$). As we discuss in detail below, we have carefully
checked that this is indeed the case for our wind solutions, and that,
for particular values of the initial velocity (inside the range
mentioned above), our solutions satisfy the critical point conditions
for a purely radiation-driven wind.

We next discuss the solutions obtained with our code {\em QWIND} for
different choices of initial parameters. First, we concentrate on a
particular `baseline' solution, in order to understand the physics of
the wind.  Then, in Section 3.3 we discuss the stability of our
solutions. In
Section 3.4 we discuss the dependence of the disc properties on the 
initial physical conditions. 
In Section 3.5 we briefly discuss a small survey of parameters in order
to show the potential of our method in studying the wind properties in a
variety of physical conditions.

A systematic analysis of all the variables,
including different density profiles, and the effect of toroidal
magnetic fields, will be the subject of a forthcoming paper.

%In the following we discuss the results of the code in two scenarios:
%1) Density and initial velocity constant for the whole wind
%(independently from the radius at which each wind element is
%launched); 2) Density and initial velocity dependent on the radius.

%The first case allows to study the dependence on the single initial
%parameters, and to put interesting constraints on them. The second,
%more realistic case allows to study under which conditions an inner
%and outer radius for the wind can be obtained as a result of the
%code, and not as external initial condition.

%%%%%%%%%%%%%%%%%%%%%%%%%%%%%%%%%%%%%%%%%%%%%
\subsection{Results: The baseline model, a case study}

We show here the results obtained from running {\em QWIND} with the
initial parameters given in Table~1.

The inner and outer radii were chosen in order to investigate the
region where a radiation-driven wind is expected, starting from an
inner radius close to the outer UV-emitting region of the accretion disc,
and studying the solutions up to an outer radius large enough to contain the
inner broad line region. 

The initial velocity was arbitrarily chosen. We remind that the important points related to this
parameter are: 
a) its value must be small compared with the escape velocity, and b) the
exact initial value should not significantly affect the final results.

A further fundamental parameter in the model is the ratio between ionizing radiation
and bolometric emission, $f_X$. Considering that the whole 0.1-100~keV spectrum 
contributes to the gas ionization, and that the wavelength range of the
intrinsic disc emission is from optical to X-rays, a typical value for this ratio is 
$f_X\sim$15\% (e.g. Elvis et al.~1994, Risaliti \& Elvis~2004). Quite different
values are however possible, due to the large dispersion of the X-ray to optical ratio
among quasars, and to its dependence on optical luminosity (e.g. Steffen et al.~2006).
We will discuss the effects of changing this parameter in the next Sections.

%%%%%%%%%%%%%%
\begin{table}
\label{baseline}
\caption{Baseline {\em QWIND} Model Parameters}
\smallskip
\begin{tabular}{|l|c|}
\hline
Parameter  & Value\\
\hline
$R_{IN}$   & 100~$R_S$ \\
$R_{OUT}$  & 800~$R_S$ \\
$v_0$   & 10$^7~cm~s^{-1}$\\ 
$\epsilon_{EDD}$& 0.5 \\ 
$M_{BH}$      & 2$\times$10$^8$~M$_\odot$ \\
$n_{IN}$   & $2\times10^8$~cm$^{-3}$\\
$\beta$ ($n_i(R)=n_{IN}\times(R_i/R_{IN})^{-\beta}$) & 0 \\
$T$        & 2$\times$10$^6$~K \\ 
$f_X$      & 0.15 \\
\hline
\end{tabular}
\end{table}
%%%%%%%%%%%%%%

%%%%%%%%%%%%%%
\begin{figure}
\includegraphics[width=8.5cm]{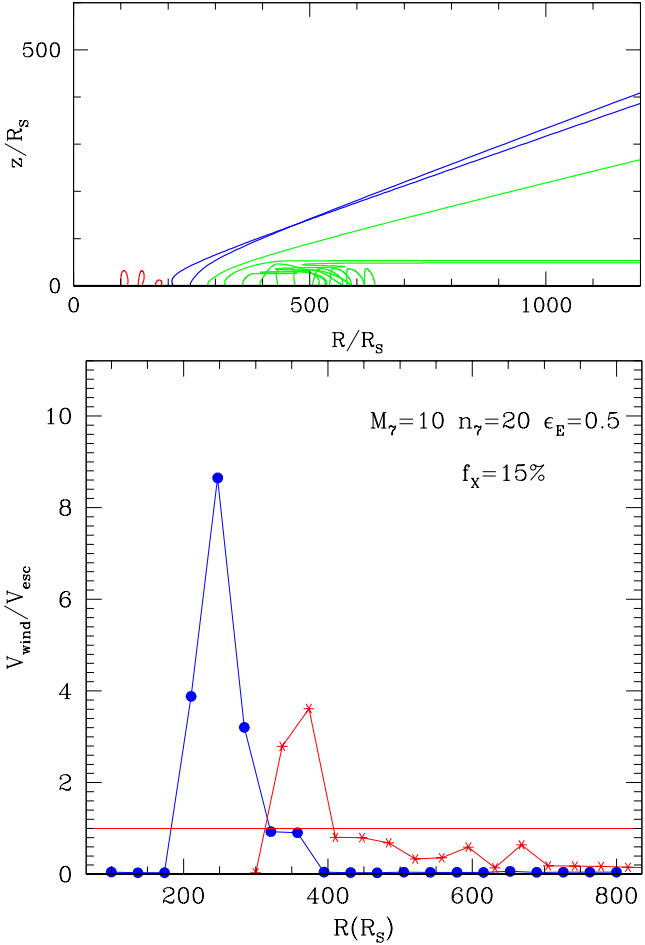}
\caption{Results of the simulations assuming a constant initial
density profile.  In this example, $R_{IN}$=100~$R_S$, $R_{OUT}$=800~$R_S$,
$n_{IN}$=2$\times$10$^8$~cm$^{-3}$, $v_0$=10$^2$~km~s$^{-1}$, $f_X$=15\%,
$\epsilon_{EDD}$=0.5, $M_7$=20. The upper panel shows the wind
geometry. In the lower panel we plot the ratio between the final
velocity of each stream line and its escape velocity, as a function of
the initial radius. We also show the final velocity versus radius
for a wind with the same initial parameters but a larger 
inner radius ($R_{IN}=300~R_G$) \label{case1}}
\end{figure}
%%%%%%%%%%%%%%

%%%%%%%%%%%%%%
\begin{figure}[h!]
\includegraphics[width=8.5cm]{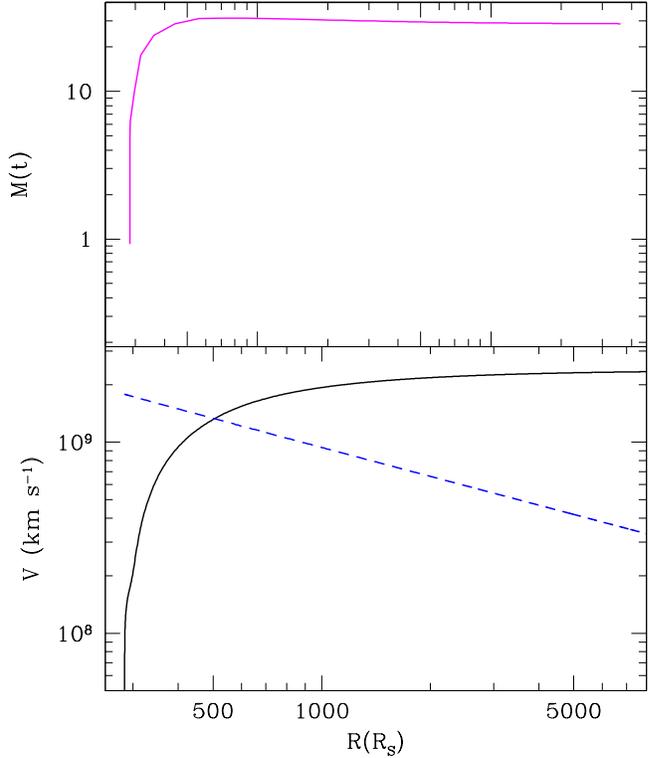} 
\caption{Force multiplier (upper plot) and velocity (lower
plots) versus radius for the fifth stream line in our case study.
The dashed line in the lower
plot represents the escape velocity as a function of the
distance from the central black hole. 
See text for details.\label{case2}}
\label{case2}
\end{figure}
%%%%%%%%%%%%%%

The main properties of the solutions are shown in Fig. \ref{case1}.
In the upper panel we show the wind lines in a plane perpendicular to
the accretion disc.  In the lower panel we show the ratio between the
final velocity and the escape velocity for each wind line.  By
definition, we have a wind when the gas velocity becomes higher than
the escape velocity, i.e. when the ratio in Fig. \ref{case1}b grows 
to values greater than 1.  In Fig. \ref{case2} we show the velocity
profile and force multiplier $M(t)$ for the fifth stream line, i.e. the one reaching
the highest final velocity.

Several interesting properties of the wind can be drawn from
Fig. \ref{case1} and \ref{case2}.

\begin{itemize}
\item {\em Initial distance}: Fig. \ref{case1}b shows that the baseline model gas
reaches escape velocity only if it is launched between 
$\sim$150~$R_S$ and $\sim$250~$R_S$.  At smaller radii, the gas is too ionized
for the radiation force to be effective. At greater radii, the UV
radiation is not enough to push the gas up to the escape velocity. The
exact values for the allowed radius range depends on the inner radius
(in our case $R_{IN}=100~R_S$).  If, for example, $R_{IN}=300~R_S$ (and
all the other initial parameters are the same), it would be impossible
to have a wind at $R=300~R_S$, because the gas is too ionized, due to
the absence of shielding from the central X-ray emission. A wind
at larger radii could however be possible, as we show in Fig.~3. {\em QWIND}
thus provides a simple explanation for the existence of a wind arising
only from a small range of distances from the center.

\item {\em Inner failed wind}: the inner stream lines form a
``failed wind'' which shields the outer stream lines from the central
X-ray radiation. This shielding is fundamental to decrease the ionization
parameter in these stream lines, so allowing an effective radiative
acceleration. This inner component resembles the ``hitchhiking gas''
of Murray \& Cheng~(1995).

%\item Velocity: the velocity profile is useful to understand the
%physical processes at work on the gas. The initial velocity is a small
%fraction of the escape velocity, then it drops because of the vertical
%component of gravity. In this phase the radiation force is not yet
%effective, because a) the UV radiation is negligible at small angles
%and b) the velocity is almost perpendicular to the radial direction,
%therefore its radial component -and its gradient- are almost null.
%When the external radiation pressure becomes important (see upper
%panels in Fig. \ref{case2}, the velocity quickly increases to its
%maximum value, and then it remains approximately constant. The final
%velocity can be as high as 10$^4$~km~s$^{-1}$.

\item {\em Radiation force}: the upper panel of
Fig. \ref{case2} shows the profile of the force multiplier $M(t)$ for
the fifth stream line in Fig. 5.  Note that the maximum value is never
higher than $\sim30$. The relatively low
values of $M(t)$ (well below the values $>100$ estimated by CAK in
hot stars) imply that only fairly large values of
\.M/\.M$_{Edd}$ ($>$0.03) can produce an escaping wind.

A more technical consequence of the estimated values of $M(t)$ is 
that the correction factor in Eq. 3 is
never important, and Eq. 2 is a good approximation of the force
multiplier (Fig. 2, i.e. $\log t >-5$). This makes possible 
an analytic analysis of the wind launching problem,
which we discuss in Section~4.

%The ``noise'' in the value of $M(t)$ is due to the dependence of the
%effective optical depth $t$ on the radial velocity gradient ($t\propto
%(dv_R/dR)^{-1}$, Eq.~1). This behavior is
%a consequence of our parametrization of $M(t)$ and on the discretization of the time 
%intervals, and has no effect on
%the observable physical quantities, provided that the time steps in
%the integration are not too long.  This is directly confirmed by the
%smoothness of the velocity profiles.

\item {\em Velocity}: Fig. \ref{case2} (lower panel) shows the velocity 
profile for the fifth stream line of our case study. The acceleration is
fast, and the escape velocity is reached at $\sim500~R_S$.
Then the velocity continues to slightly increase. 
In this phase the radiation force is still effective
in supporting the wind against gravity. Indeed, the ``effective'' Eddington
ratio, which depends on the actual one, $\epsilon_{EDD}$, the force multiplier,
and the inclination angle $\theta$ of the disc as seen from the wind line, is
$\epsilon_{EDD}\times M(t)\times \cos\theta>1$ up to large radii (R$\sim5000~R_S$).

\item {\em Geometry}: Fig. \ref{case1}a shows that the gas rises
vertically for $\sim 50~R_S$, and then bends toward a radial
direction. Whether the gas falls down or maintains this direction
depends on the effectiveness of the radiation force (see below).  The
angle of the wind above the disc is approximately 20 deg, giving a
substantial covering factor of 35\%.  We note that the actual
value of the covering angle is somewhat uncertain because the lines
are treated as independent while, in reality, when two lines
intersect, the gas pressure from the inner line pushes the gas on the
outer line outwards, altering the final covering factor.

\end{itemize}

%%%%%%%%%%%%%%%%%%%%%%%%%%%%%%%%%%%%%%%
\subsection{Check of initial conditions}

In order to test the reliability of our initial conditions,
in particular regarding the initial values of the vertical velocity and
the height above the disc, we performed two  checks:
\begin{enumerate}
\item We varied the initial height from $H=5~R_S$ to $H=15~R_S$, and
the initial velocity from 100 to 300~km~s$^{-1}$.  
No significant change in
the final properties of the wind was found.
\item We followed the wind streamlines backward in time down to the
critical point (estimating this location as in Murray et al.~1995) and
required that our solutions satisfy the regularity and stability
conditions at the critical point (CAK). We iteratively varied the
initial velocity until a consistent solution was found.  A solution
was found with a value of the initial velocity within our range of
study ($v_0$=100-300~km~s$^{-1}$).
\end{enumerate}

The solution provided by {\em QWIND} thus represents the supersonic part of a purely
radiation-driven wind (i.e. the case where, even in the subcritical
phase, the radiation force is the only external force present in
addition to gravity) and is in all respects equivalent to that found
by Murray et al.~1995. The existence of such solutions within the
range of initial velocities we are exploring shows that our approach
is well justified, at least in the pure radiation-driven scenario.

%%%%%%%%%%%%%%%%%%%%%%%%%%%%%%%%%%%%%%
\subsection{Dependence on the individual parameters}

In order to explore the dependence of our solutions on the initial
parameters, we proceed in two steps: First, we analyze several cases
close to the baseline model, changing the density, the X-ray to UV ratio, 
the Eddington ratio, and the balck hole mass. Then, in the next Subsection 
we show the results of a survey of
the $\epsilon_{EDD}-M_7$ space for a few choices of the other initial
parameters to show the potential of our method in exploring the 
initial conditions parameter space.
%%%%%%%%%%%%%%%
\begin{figure}[h!]
\includegraphics[width=8.0cm]{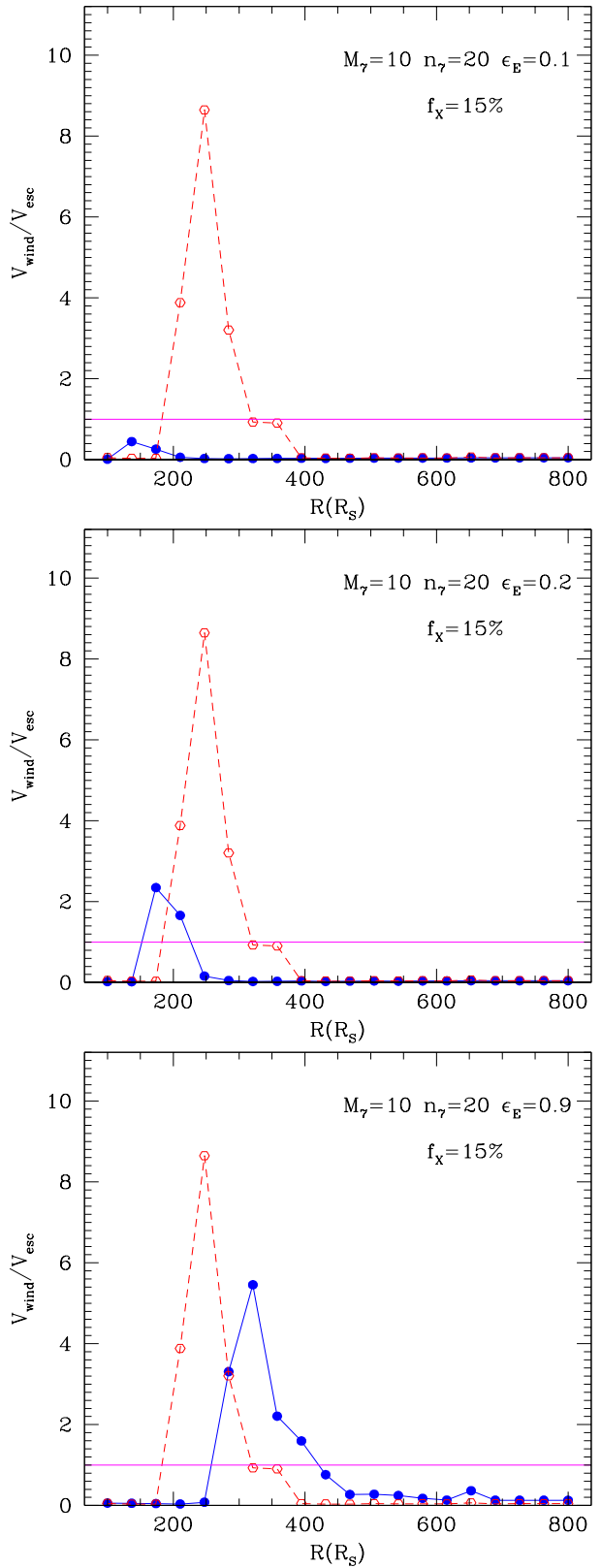} \caption{Ratio between the gas final velocity and
its escape velocity for several values of the Eddington ratio, $\epsilon_{EDD}$.
The red dashed line reproduces the same profile for the ``case study'' shown 
in Fig.~3.
\label{par1}}
\end{figure}
%%%%%%%%%%%%%%%

%\item In the first column we show the dependence on the {\em initial
%velocity}, $v_8$.  The most interesting result is the existence of a
%threshold value for the initial velocity (10$^{6.5}$cm~s$^{-1}$),
%below which it is impossible to have a wind. The physical reason for
%this apparent discontinuity is in the nature of the radiation force
%term, given by Eqs. 1 and 2. In order for the radiation force to be
%effective, a significant radial velocity gradient is needed. As we
%pointed out above, when this condition is reached, the acceleration
%produced by the radiation force itself increases this gradient, which
%in turn makes the radiation force greater and so on.  Therefore, there
%is a critical starting condition that must be reached.  This condition
%depends both on the actual velocity of a given gas element, and on the
%height it can reach before starting to falling back into the disc
%(this height is important both for the incident UV radiation, and for
%the projection of the velocity along the radial direction, both
%dependent on the angle formed by the disc and the radius relative to
%the gas element).  If this critical condition is not reached, the
%force multiplier never becomes significantly greater that 1, and no
%wind can be launched.

First, starting from the baseline model (Table~\ref{baseline}), we
systematically changed $\epsilon_{EDD}$, $f_x$, $n_8$, $M_{BH}$, one at a time:
In Fig. \ref{par1}-\ref{par4} we show plots analogous to that in
Fig. \ref{case1}b, i.e. the ratio between the final wind velocity and
the escape velocity, $v/v_{ESC}$, against the initial wind radius.\footnote{In these
figures, and in the next Sections, we adopt the convention $n_7$=$n$/(10$^7$~cm$^{-3}$),
$M_8$=$M_{BH}$/(10$^8$~$M_\odot$).}
From these results several additional indicators of the physical
processes dominating the wind can be obtained, based on the dependence
of $v/v_{ESC}$ on the:

\begin{itemize}

\item {\em Eddington ratio $\epsilon_{EDD}$} (Fig. \ref{par1}): A too
small value (panel A, $\epsilon_{EDD}=0.1)$ implies that the
radiation is inadequate to accelerate the gas to escape velocity. With
a higher value (panel B, $\epsilon_{EDD}=0.2$, a stream line is
able to slightly exceed the escape velocity, thus forming a wind.
Finally, for much higher values (panel C, $\epsilon_{EDD}=0.9$, 
to be compared with our case study with $\epsilon_{EDD}=0.5$)
a wind is effectively launched, but at larger radii than in the previous cases.
This is due to the higher X-ray luminosity, which  over-ionizes
the gas up to larger distances from the center.

It is interesting to note that the maximum velocity reached by the 
wind in the case $\epsilon_{EDD}$=0.9 
is smaller than in the baseline case of Fig.~\ref{case1}, despite
the higher Eddington ratio.
This is due to the reduction of the UV radiation pressure on the
stream lines with the ``right'' ionization state, due to the higher distance.
This result shows in a simple way that, due to the different 
dependence of the the physical conditions of the wind elements on the
initial parameters, a higher luminosity does not automatically imply a 
faster wind. In our case the intuitive increase of the wind velocity
is observed increasing the Eddington ratio from 0.2 to 0.5, but not
with the further increase to 0.9. 
\item {\em Density of the gas} (Fig. \ref{par2}): 
At small
values of the density (left plot, $n_7=3$, to be compared with $n_7=20$
in the standard case discussed in the previous Section), the gas is
overionized even at large radii, and it is therefore impossible to
launch a wind.  Increasing the density ($n_7=10$, right plot) the
ionization parameter decreases, making it easier to push the gas
effectively through the line-absorption driven force. An even higher density
shifts the wind towards inner radii, because the over-ionization region is limited to 
a smaller region. 
\begin{figure}[h!]
\includegraphics[width=8.0cm]{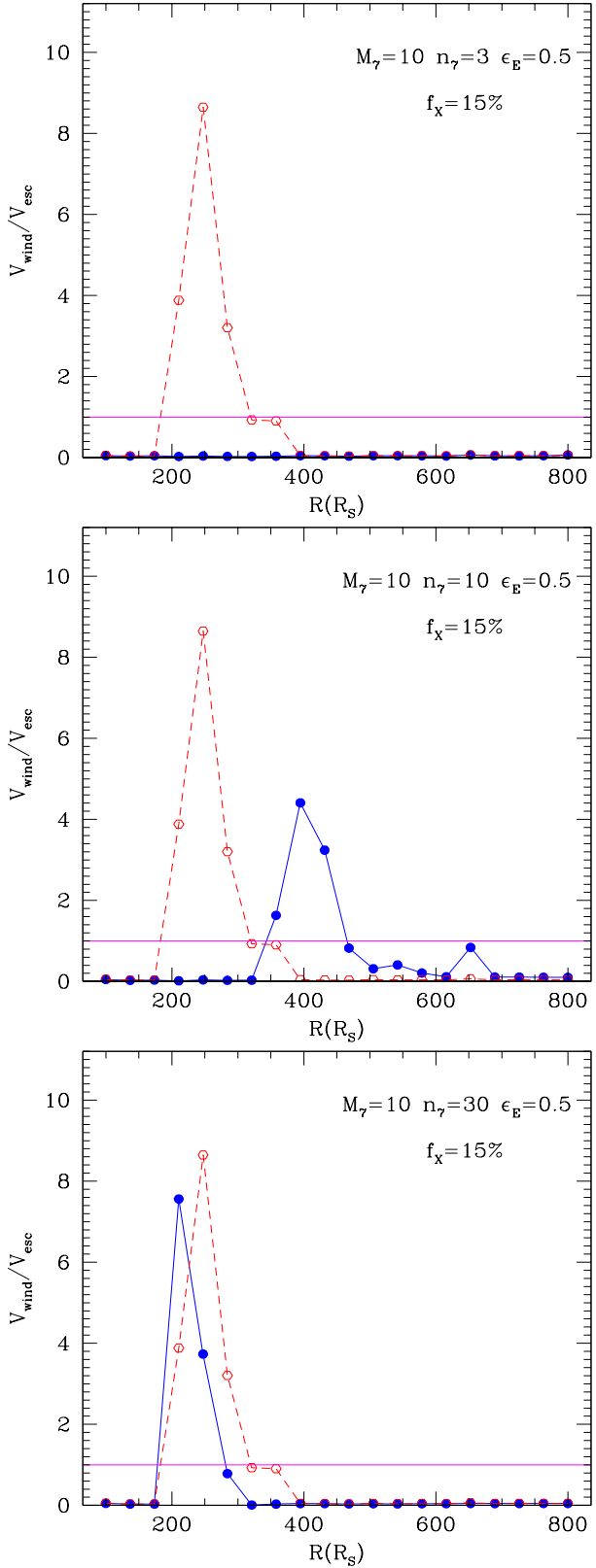} \caption{Ratio between the gas final velocity and
its escape velocity for several values of the initial gas density, $n_{IN}$.
The red dashed line reproduces the same profile for the ``case study'' shown 
in Fig.~3.
\label{par2}}
\end{figure}
\item {\em The ratio $f_X$} between X-ray and disc radiation (Fig. \ref{par3}): 
Too high a value of $f_X$ (in our example, $f_X=$25\%, corresponding to
a 2~keV to 2500\AA~slope of $\alpha_{OX}=1.3$) causes a
higher ionization, and make it more difficult to accelerate the gas up
to the escape velocity.  Smaller values ($f_X=$5\%, $f_X=$10\%, corresponding
to $\alpha_{OX}=1.6$ and $\alpha_{OX}=1.5$, respectively) mean a
smaller ionizing continuum, and therefore a lower ionization factor
for the inner gas of the wind, which can be accelerated up to the
escape velocity. This effect was first noted by Murray \& Chiang
(1995). 
\begin{figure}[h!]
\includegraphics[width=8.0cm]{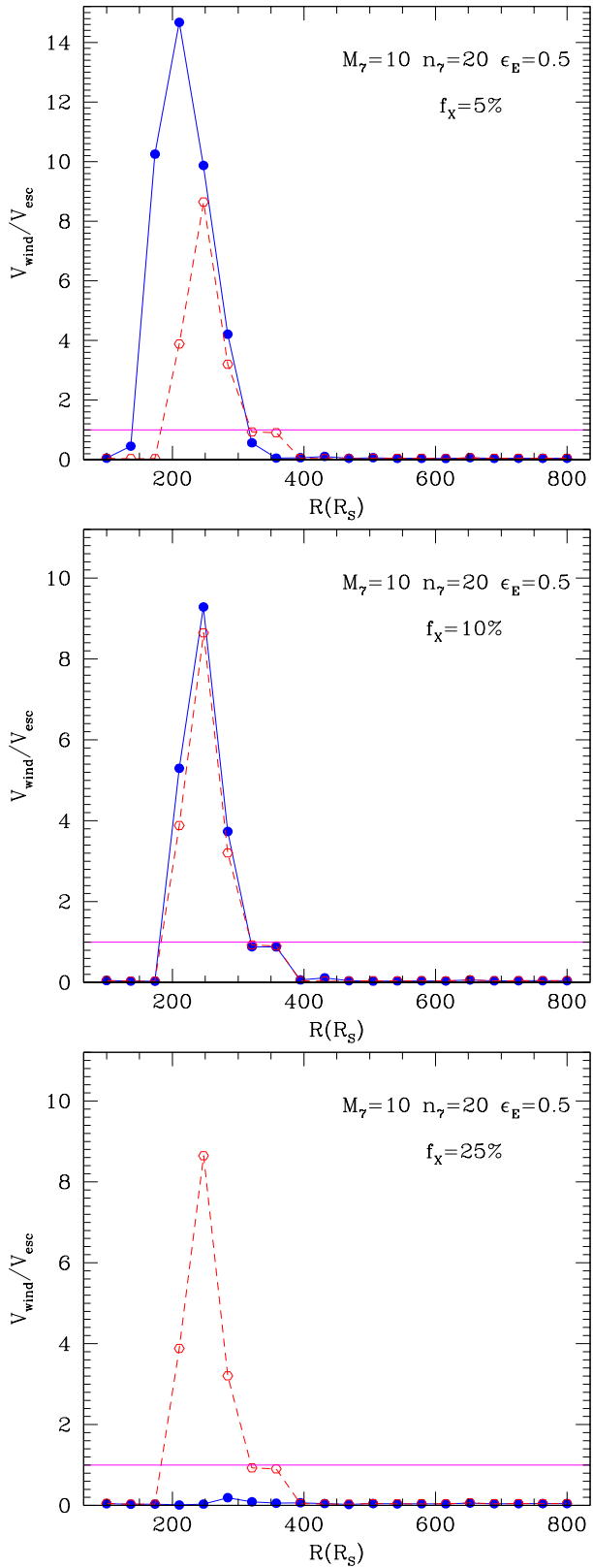} \caption{Ratio between the gas final velocity and
its escape velocity for several values of the ratio $f_X$ between the X-ray and UV flux.
The red dashed line reproduces the same profile for the ``case study'' shown 
in Fig.~3.
\label{par3}}
\end{figure}
\item {\em Black hole mass} (Fig. \ref{par4}):
The wind properties strongly depend on the black hole mass. The dependence is
quite complex, since the different physical parameters (ionization state, UV flux, disc temperature) scale in different ways with the black hole mass. Varying the black hole 
mass and leaving the other parameters as in our case study, we see that the higher the mass,
 the closer in is the wind. For the highest mass value ($M_{BH}=10^9~M_\odot$) no wind is launched with the adopted choice of the initial parameters. The main physical driver of this behaviour is the ionization parameter, $U\propto L/R^2$. Since $L\propto M_{BH}$ (for a fixed $\epsilon_{EDD}$), {\bf and the distance in physical units is $R\propto$M$_{BH}$ (for a fixed value in units of R$_S$), the ionization parameter at a given distance in units of $R_S$ decreases with $M_{BH}$. The distance (in units of $R_S$) 
at which we have the ``right'' balance between
ionization state and UV irradiation therefore decreases with increasing black hole mass, reaching
values lower than our inner radius for $M_{BH}=10^9~M_\odot$. In this case, it may be possible to
obtain a wind from radii R$<$R$_{IN}$. This scenario will be studied in a wider analysis of the parameter space which  will be presented in a forthcoming paper.
 Here we only note that this result does not imply that a wind is impossible 
at high masses: an example of wind from a 10$^9$~$M_{\odot}$ black hole is presented in the next 
Subsection, where we show that a different choice of the initial density leads to higher ionization
 parameters, shifting the "wind zone" towards higher radii. In particular, we will show that an 
initial value of the density $N_7$=1 (i.e. 20 times smaller than our baseline case) implies that a
 wind can be launched {\em only} with black hole masses of the order of 10$^9$~M$_\odot$.}
%This systematic study will be the subject of a forthcoming paper.
\end{itemize}
\begin{figure}[h!]
\includegraphics[width=8.0cm]{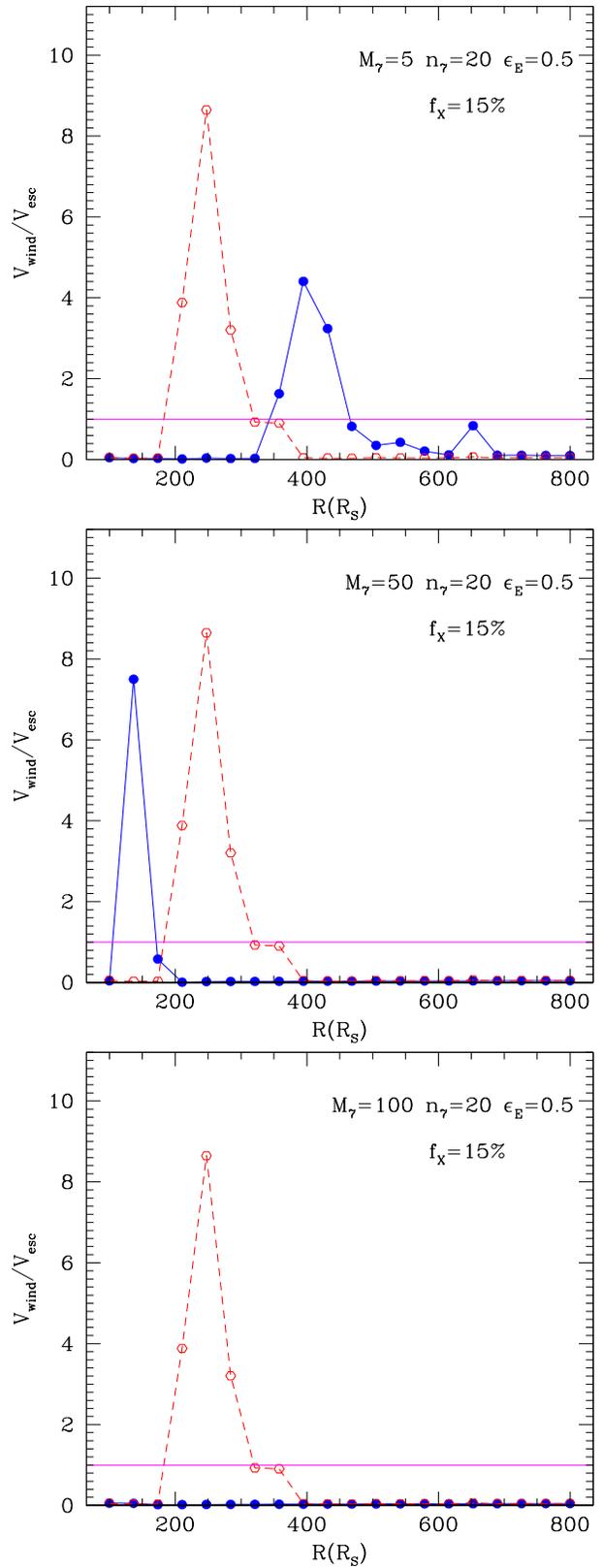} \caption{Ratio between the gas final velocity and
its escape velocity for several values of the central black hole mass, $M_{BH}$.
The red dashed line reproduces the same profile for the ``case study'' shown 
in Fig.~3.
\label{par4}}
\end{figure}
\subsection{Parameter survey}

\begin{figure}[h!]
\includegraphics[width=8.0cm]{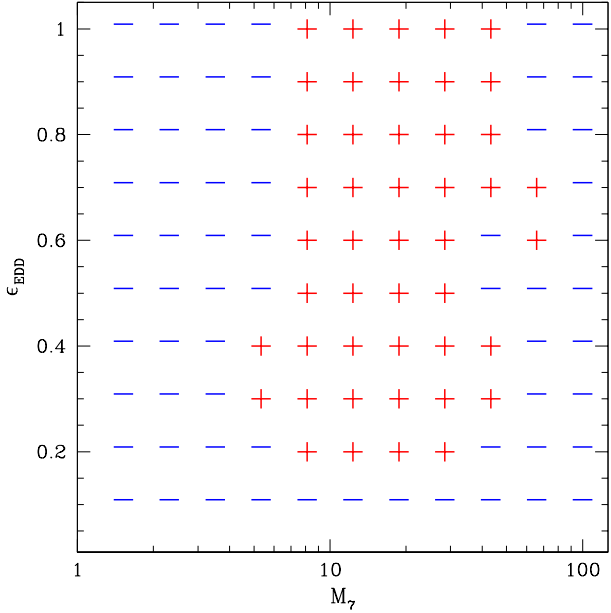} \caption{Results of a survey of the $\epsilon_{EDD}-M_{BH}$ plane, with the other initial parameters equal to those of the baseline model. The ``$+$'' indicate that a wind is launched.
\label{survey1}}
\end{figure}

In order to test the ability of {\em QWIND} to perform complete surveys of the parameter space,
we run a grid of 11x10 models, adopting the parameters of our baseline study, and varying the black hole mass and the Eddington ratio in the range 10$^7$-10$^9$~$M_\odot$ and 0.1-1, respectively. 
The result is shown in Fig.~\ref{survey1}.
The allowed region for the wind launching extends down to $\epsilon_{EDD}\sim0.2$, showing that a radiation wind is possible with luminosity much lower than the Eddington limit
(as also shown in more detail for a single case in Fig.~\ref{par1}b). Clearly, in these cases
the enhancement of the radiation force due to line absorption plays a fundamental role.

\begin{figure}[h!]
\includegraphics[width=8.0cm]{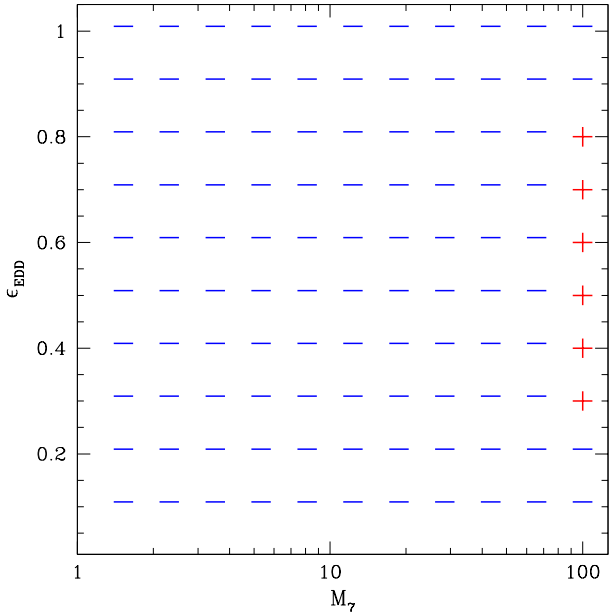} \caption{Results of a survey of the $\epsilon_{EDD}-M_{BH}$ plane, with an initial gas density $n_{IN}=10^{7}$~cm$^{-3}$, 1/20 of the baseline value. The ``$+$'' indicate that a wind is launched.
\label{survey2}}
\end{figure}

As a further example of the application of {\em QWIND} we also tested the scenario of a hot, low density wind
with a less
substantial mass outflow. To test this, we run a {\em QWIND} grid analogous to the one described above, but with an initial density 20 times smaller ($n_{IN}=10^7$~cm$^{-3}$). The results are shown in Fig.~\ref{survey2}. In general, a too low density implies a too high ionization, and the wind cannot be launched. The situation is different only at very high masses, where the distances from the central source are large enough to provide the requested balance between ionization parameter and intensity of the UV source.\\

\section{Analytic approximations}

Since our main aim is to understand which physical conditions
allow wind acceleration, we can search for further
simplifications which allow an even simpler treatment, based on an
analytic model.

In this Section we will make three extra approximations to the dynamical
equations of the wind, which allow the equation of motion to be
integrated analytically, so that the final velocity can be calculated
directly. The approximations are:

\begin{enumerate}
\item an  energy source with a constant inclination with respect to
the wind. 
For the X-rays, this is the same as assumed in {\em QWIND} (spherically symmetric 
emission). For the UV, this is equivalent to the assumption of a fixed
inclination angle of the wind stream lines. 
\item neglect the gravitational term in the dynamic equation, as once
the line-driven acceleration mechanism becomes effective, it likely
dominates the gravitational term by a factor of 10 or more (see
Section 4.3).
\item line driving is well reproduced by the simple equation
$M(t)=K\times t^{-\alpha}$, which is valid only when the wind is not
overionized, but we require a low ionization parameter as a condition
for the acceleration of the wind (see below).
\end{enumerate}

Our aim is not to study the details of the wind solutions, but rather
to explore the parameter space in order to understand which are the
initial conditions required in order to have accelerations up to several
thousand km~s$^{-1}$.  The analytic treatment lets us search for the
existence of wind solutions by finding regions of parameter space that
satisfy two conditions: (1) that the final velocity exceeds the escape
velocity; and (2) that the gas must not be overionized.

Each of these three approximations makes wind solutions more likely:
(1) the spherical approximation implies that the UV radiation is not
decreased by the geometrical projection factor of the disc, (2)
neglecting the gravitational term obviously favors fast outflows, and
(3) the force multiplier assumption maximizes the effect of
line-driving.  As a consequence we will obtain wind solutions in
larger regions of the parameter space than found using the more
precise numerical approach described in Section~3.  We will
show that in practice the analytically allowed parameter space is
close enough to the results of the numerical code to make the
analytical results useful in delimiting large regions of the
multi-dimensional initial parameter space where it is {\em not}
possible to accelerate a wind.

We discuss the third assumption, the conditions on the ionization
state, in the next Subsection.  The integration of the equation of
motion, and condition on the final velocity are discussed in Section
4.2. We then show the results for a limited set of initial parameters,
and finally we compare our results with those obtained with {\em QWIND}.

%%%%%%%%%%%%%%%%%%%%%%%%%%%%%%%%%%%%%%%%%%%%%%
\subsection{Relation between $M(t)$ and $\xi$}

Here we show that the equation of the force multiplier, M(t) can be
simplified into two regimes, depending on the ionization factor: $\xi
< 100: M(t)=K t^{-\alpha}, \xi>100: M(t)=0$.  We then discuss the
equation describing the motion of a gas element inside the wind.

In order to estimate the range of values of $t$ for which the
correction factor of M(t) becomes important, and therefore M(t)
saturates to its asymptotic limit, we expand Eq. 3 for small $t$,
obtaining:
%%%%%%%%%%%%%%%
\begin{equation}
M(t)\sim K(1-\alpha)\eta^\alpha_{MAX}(1-\alpha\eta_{MAX}t)
\label{eqmt4}
\end{equation}
This shows that in order to have
$M(t)\rightarrow(1-\alpha)\eta_{MAX}^\alpha$ the effective optical
depth must be
\begin{equation}
t<<\frac{1}{\alpha\eta_{MAX}}
\end{equation}
%%%%%%%%%%%%%%%
From Fig.~\ref{stevens1} , for $\xi<100$ this implies
t$<10^{-6}$. This condition can also be easily obtained from
Fig.~\ref{stevens2}.  For an outflowing wind reaching a radial
velocity of $\sim 10^4$~km~s$^{-1}$ in a distance of
10$^{16}-10^{17}$~cm, typical AGN values, we have an average $\frac{dv_r}{dr}
\sim 10^{-7}-10^{-8}$~s$^{-1}$ and so, from Eq. 1, $t\sim 7\times
10^{-3} T_6^{0.5} n_8$, where $T=10^6T_6$ K and $n=10^8n_8$ cm$^{-3}$.
It is easily seen from Fig.~\ref{stevens1} that if the ionization
factor $\xi$ is lower than 100, M(t) is never close to the saturation
regime for reasonable values of the density and temperature, and
$M(t)=K\times t^{-\alpha}$ is a good approximation.  $K$ is constant
within a factor $\sim$2 in the range of interest (Fig.~1).

If $\xi>100$ the situation is completely different since it is obvious
from Fig.~\ref{stevens1} that the line absorption becomes negligible
compared with electron scattering and in this condition no wind can be
launched unless the luminosity is super-Eddington.  This is why the
wind models of Murray \& Chiang (1997) require additional,
``hitchhiking '' gas, to shield the gas that was to be accelerated,
reducing $\xi$.  Therefore, in the following we shall assume $\xi <
100$ in studying the dependence of the wind from the other parameters.

Now let us consider a gas element in the outflow at radial distance R from
the center. The gas element is subject to gravitational force, $F_G=GM_{BH}
m_p R^{-2}$, and to the radiation force, $F_r=L(4\pi R^2
c)^{-1}\sigma_T (M(t)+1)$.  The force multiplier $M(t)$ is given by
Eq. 3, and depends on the density profile of the gas making up the
wind.  We assume a profile $n_i(R)=n_{IN}\times (R/R_{IN})^{-\beta}$,
as in our numeric code.
%that the initial density of this gas scales with the
%density of the accretion disc. We therefore use a SS density profile
%
%\footnote{The actual theoretical result for a SS disc is $n\propto
%{-15/8}$ (SS). Here we approximate to $n\propto R^{-2}$, in order to
%write the equations in a simpler form. The difference is negligible in
% cases.}
%
%, $n\propto R^{-2}$.

The flux at a given gas element inside the wind is given by
$F L_0/(4\pi R^2)\times~A(R)$, where $L_0$ is the intrinsic disc luminosity, 
$F$ is the geometrical factor due the disc inclination (which is assumed to be
constant in
this analytical treatment) and
$A(R)$ takes into account the absorption between the gas element
under consideration and the luminosity source.  While a detailed
treatment of the absorption term would require a complete solution of
hydrodynamical and radiative transfer equations, we can make some
useful approximations:
\begin{itemize}
\item that the inner part of the gas is able to reach high enough
above the disc in order to shield the gas farther out. This assumption
is supported by the results of {\em QWIND} (\S 3), and has the same
effect as 
the assumption of hitchhiking gas by Murray \& Chiang (1995).
\item that each ring of gas with thickness $\Delta R$ contributes to
radiation absorption with a column density of $\sim n\Delta R$. This
is obviously a lower limit, since the wind is not perpendicular to the
central radiation. However, the correction is small and can be
neglected.
\end{itemize}

With these two assumptions, the absorption term can be written as:
%%%%%%
\begin{equation}
A(R)=e^{-\sigma_T\int^R_{R_{IN}} n(R') dR'}
\end{equation}

where $\sigma_T$ is the Thomson cross section.  Assuming the density
profile as in our numeric code, $n(R)=n_{IN}\times(R/R_{IN})^{-\beta}$,
we have:

%$n(R=R_{IN})=10^8$~cm$^{-3}$, we have:
\begin{equation}
A(R)=e^{-\frac{0.02}{\beta-1}~n_8 R_{100} M_7[1-(\frac{R}{R_{IN}})^{1-\beta}]}.
\label{eqnh}
\end{equation}
%%%%%%

We can now write the equation of motion for the gas, simply requiring
that the force on a gas element is equal to the difference between the
external radiation force and the gravitational force.  As in the {\em QWIND}
code (\S 3), we neglect the internal pressure term, since we want to
discuss the motion of the ion in the supersonic part of the wind,
where gas pressure cannot affect the dynamics significantly.  Furthermore, we assume that only
radial forces are present. This implies the conservation of angular
momentum, and a decoupling of the radial and tangential equations. The
radial equation of motion can then be written, putting
$\epsilon_{EDD}$ for the Eddington ratio and $R_{IN}$ for the initial
radius, as:

\begin{equation}
v_R\frac{dv_R}{dR}=\Phi_{RAD}+\Phi_{GRAV}
\end{equation}

where the left hand side gives the radial acceleration, while the
right hand side is made up of two terms: 
\begin{enumerate}
\item the line-enhanced radiation force, given by the flux multiplied by the force
multiplier M(t), and by a geometrical factor, $F(\phi)$, dependent on
the inclination angle $\phi$ of the disc with respect to the gas
element:
\begin{equation}
\Phi_{RAD}=\frac{GM}{R^2}\epsilon_{EDD} A(R) F(\phi)K (n v_{TH}\sigma_T \|
\frac{dv_R}{dR}\|^{-1})^{-\alpha} \\
\end{equation}

\item the gravitational force, corrected by a factor
$(1-\frac{R_i}{R})$, obtained from angular momentum conservation,
assuming that the gas element moves with the Keplerian velocity at the
initial radius $R_i$, and decreased by the continuous-radiation force,
due to Thomson scattering):
\begin{equation}
\Phi_{GRAV}=-\frac{GM}{R^2}(1-\frac{R_i}{R})(1-\epsilon_{EDD}F(\phi) A(R))
\label{eqmotion}
\end{equation}

\end{enumerate}

A further fundamental condition is given by the mass conservation
condition, $\dot{M}=4\pi R^2 n v={\rm const}$, which implies that
the only stable solutions are those through the critical point, as
first discussed by CAK75 and, in a context similar to ours, by Murray
\& Chiang (1995). (See also Lamers \& Cassinelli, 1999, for a textbook
treatment.)

In general, Eq.~\ref{eqmotion} can only be solved using numerical
codes. However, in our case we are interested in some special
situations. In particular, the wind solution we are seeking is
characterized by a sudden and strong radial acceleration (as demonstrated by
the results of {\em QWIND} shown in Section~3),
which changes the gas motion from slow and quasi-vertical to radial
and fast (with velocity higher than the escape velocity).  We can
assume that in this accelerating phase the external force is dominant
with respect to the gravitational term, which can be neglected. We
also require that the final wind velocity, $v_F$, exceeds the escape
velocity, $v_{ESC}=\sqrt{2GMR^{-1}}$, within a distance from
the center of a few $10^3~R_S$. We will then check whether our
solutions fulfill the conditions described above. This consistency
check will only show whether our wind solutions are acceptable, but
other solutions are also possible.  For example, we expect that for a
large set of initial conditions, the gas, after initially rising will
fall back onto the disc (in these cases the gravitational term is
obviously not negligible).

Neglecting the gravitational term, and requiring mass conservation, we obtain:
\begin{equation}
\label{eqmotion2}
R^2v_R\frac{dv_R}{dR}=\eta[R^2v_R\frac{dv_R}{dR}]^\alpha
%-GM(1-\frac{R_i}{R})
\end{equation}
where
\begin{equation}
\eta=(v_{TH}\sigma_T v_i n_i R_i^2)^{-\alpha} K F(\phi)A(R)GM\epsilon_{EDD}
\end{equation}
Since we are studying solutions with rapid radial accelerations, which quickly make 
the gas motion radial, we can assume that the geometrical factor is constant ($F(\phi)=F_0$).
{\bf Furthermore, we neglect the dependence of $A(R)$ on the distance $R$,
assuming A(R)=A(1.2R$_i$). This approximation is reasonable, considering that
our simulations with {\em QWIND} (Section~3) show that when a wind is
successfully launched, the escape velocity is reached at a distance from
the center $R_{ESC}$ only slightly 
larger (by on average $\sim$20\%, from Figs.5-8) 
than the initial distance of the stream line, $R_{i}$.} 
With these assumptions, Eq.~\ref{eqmotion2} can be easily integrated between an initial
radius, $R_i$, and a final radius, $R_F$. 
The condition $v_F>v_E$ can then be written as:
\begin{equation}
\eta^{\frac{1}{1-\alpha}}r_i^{-1}[1-\frac{R_i}{R_F}]>GMR_F^{-1}
\end{equation}
Finally, we adopt  the initial density
profile discussed above, and we assume $R_F$=2$R_{i}$.
A choice of $R_F$ is needed in our treatment, because having neglected
the gravitational term implies that the above equation is always satisfied 
for $R_F \rightarrow \infty$.
We assume $R_F$=2$R_i$, which, based on the above considerations on $R_{ESC}$,
is a rather conservative assumption (in all the cases analyzed with {\em QWIND}
we found $R_{ESC}$$<$2$R_i$), 
% However, this assumption is valid in the
%wind phase dominated by the radiation pressure term.
% Indeed, in all cases
%we analyzed, $R_F<2 R_{i}$. 
%Therefore, our assumption is conservative,
in the sense of allowing a larger parameter space for wind launching than in our analytic
treatment. This is in agreement with our general approach, as discussed at
the beginning of the present Section.

{\bf We can then rewrite the above equation as:
\begin{equation}
[v_{TH}\sigma_T v_I  n_I R_i^2 (GM)^{-1})]^{-\alpha} k F_0 A \epsilon{EDD} >1
\end{equation}
Finally, adopting a proper parametrization for the physical quantities and
substituting the numerical values for the constants, and assuming
$\alpha$=0.6, we obtain:
}
%with proper parameterizations for the physical quantities involved. After 
%straightforward algebra, and using the value $\alpha=0.6$ we obtain:
\begin{equation}
\label{eqtheory1}
\epsilon_{EDD}>1.1\times10^{-3} A^{-1}F_0^{-1}[\sqrt{T_7}v_7n_8R_{i,100}^2M_7^{-1}]^{0.6}  
%1.1\times10^{-5} M_7 n_8 v_7 R_{i,100}^2 \sqrt{T_7} (\epsilon_{EDD} A(R)F(\phi))^{-\frac{5}{3}}>1
\end{equation}
This condition, when combined with the requirement that the gas not be
over-ionized (\S 4.2), will allow us to put constraints for the
existence of wind-like solutions in a black hole mass - Eddington
ratio plane, as we will show in Section 4.3.

%%%%%%%%%%%%%%%%%%%%%%%%%%%%%%%%%%%%%%%%%%
\subsection{Effects of X-ray absorption}

If the inner part of the wind is too close to the X-ray source, an
over-ionization problem arises.  An approximate treatment can be
obtained by considering the X-ray cross-section of the wind gas as in
P00.  Using the approximation described in Section 3, a fundamental
requirement for a radiation-driven wind to be launched is that
{\bf $\xi$=L/(n(R)R$^2$)$<$10$^5$, where L is the ionizing radiation.  
Adopting the same parametrizations as in the previous Subsection,
the above condition can be written as:}  
\begin{equation}
\label{eqlowxi}
1.5\times10^8 f_X n_8^{-1}R_{100}^{-2}M_7^{-1}A(R)\epsilon_{EDD}
<10^5
\end{equation}
where $A(R)$ is the fraction of non-absorbed X-ray radiation, and is
given by Eq.\ref{eqnh}, the cross section being the same (for
$\xi>10^5$) as for the UV disc radiation. $f_X$ is the fraction of the
bolometric luminosity emitted in the X-rays. In the work of 
Stevens \& Kallman (1986) the objects considered as sources of
radiation-driven winds were X-ray binaries, with a rather hard
spectrum. As a consequence, only photons with E$>$1~keV were taken
into account in their estimate of the ionization parameter.

In our case, the typical intrinsic X-ray spectrum of an AGN is dominated
by soft emission in the 0.1-1 keV band, which gives a significant -if
not dominant- contribution to the ionization parameter.
Therefore, we calculate the ionization parameter using the luminosity 
above 0.1~keV. As a standard value we assume that of Elvis et al. (1994),
where the 0.1-100~keV luminosity is $\sim15$\% of the total
(i.e. optical-UV-X-ray) emission of an AGN.

Assuming the density profile of the SS disc $n(R)=n_{IN}\times(R/R_{IN})^{-\beta}$
equation~\ref{eqlowxi} can be finally rewritten as:
%{\sc re-ordered to be the same as Eq.13}
\begin{equation}
\epsilon_{EDD}<0.0044 (\frac{f_X}{0.15})^{-1} M_7 n_8(\frac{R}{R_{IN}})^{-\beta} A(R)^{-1} R_{100}^2
\label{eqtheory2}
\end{equation}
%where 
%$n_7$ is the density at the wind inner radius. 
%{\sc should be $n_{in}$?}

%%%%%%%%%%%%%%%%%%%%%%%%%%%%%%%%%
\subsection{Analytical Results}

The constraints from Eq.\ref{eqtheory1} and Eq.\ref{eqtheory2} delimit
the region of the black hole mass, Eddington ratio ($M_{BH},
\epsilon_{EDD}$) parameter space in which an outflowing wind can
arise.  

We can use Eqs.\ref{eqtheory1} and \ref{eqtheory2} to plot the allowed
wind-launching regions in the $\epsilon_{EDD}-M_{BH}$ plane.
Since Eq.~\ref{eqtheory2} depends on the initial radius of each streamline,
for a given set of initial parameters (wind inner radius $R_{IN}$,
density at the inner radius $R_{IN}$, gas temperature $T$ and X-ray to
optical/UV ratio $f_X$) we compute the allowed region for each single 
streamline, and then we plot the convolution of all these regions.
In this way, the meaning of the ``analytic parameter surveys'' are
analogous to those plotted in Figs.~\ref{survey1} and~\ref{survey2}.

We assume a baseline set of initial parameters (Table~\ref{baseline}),
as used in Eqs.\ref{eqtheory1} and \ref{eqtheory2}: $f_X=$15\%,
$T=2\times10^6$~K, $n_{IN}=10^8$~cm$^{-3}$ at $R_{IN}=100~R_S$.  These
parameter values are those required for the gas responsible for the
observed UV and X-ray `Warm absorber' absorption lines (Nicastro et
al. 1999, Netzer et al. 2002, Krongold et al. 2003) in quasar
spectra. They are also the physical conditions needed for a gas
confining the BEL clouds (which have $n\sim 10^{10}$~cm$^{-3}$ and
$T\sim 10^4$~K, Osterbrock 1989) as noted by Turner et al. (1994) and
Elvis (2000).
%%%%%%%%%%%%%%%
\begin{figure}[h!]
\includegraphics[width=8.5cm]{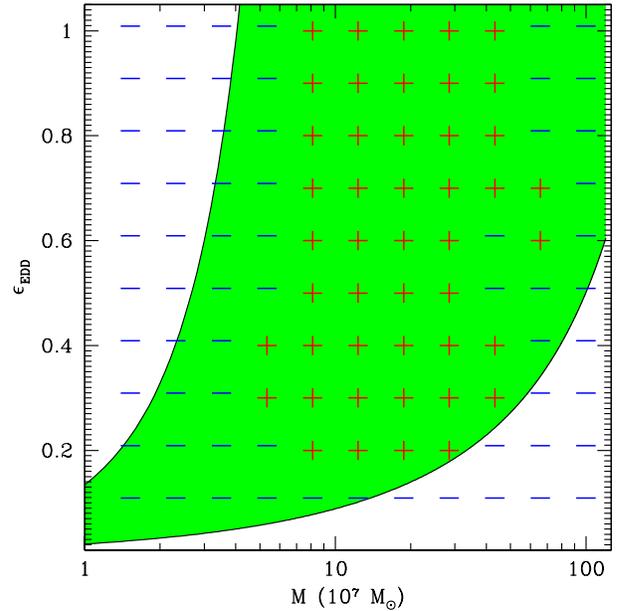} 
\caption{Results of our numerical analysis for our baseline set of
parameters. The green (shaded) region indicates where a wind
can be launched.  The analytic estimate is compared with the 
numerical result obtained with a {\em QWIND} survey. 
\label{theory1}}
\end{figure}
%%%%%%%%%%%%%%%

Figure \ref{theory1} shows the allowed ($M_{BH}, \epsilon_{EDD}$)
parameter space for the baseline parameters, superimposed to
the results of {\em QWIND} for the same set of parameters (already shown
in Fig.~\ref{survey1}.
The two interesting results emerging from this plot are:\\
1) The 
allowed region based on our analytic approximation is larger than
that obtained with the more detailed approach based on  {\em QWIND}. This
confirms that our approximations do not miss any possible wind solution,
in agreement with our expectations (Section~4.1).\\
2)  The results of the analytic approximations
are useful to exclude a  significant part of the parameter space
where a wind cannot exist. This approach can then be used for
a genaral exploration of the parameter space, leaving the more
time-consuming numerical approach for a smaller set of parameters.
This was not obvious {\em a priori}, since a too strong approximation
could have lead to a too large (and therefore useless) allowed region.

While a complete investigation of the parameter space is not the
purpose of the present work, it is useful to show the results of
the analytic approach for a small set of initial parameters, as 
examples of the possible uses of this technique.

\begin{figure*}
\centerline{\includegraphics[width=16.5cm]{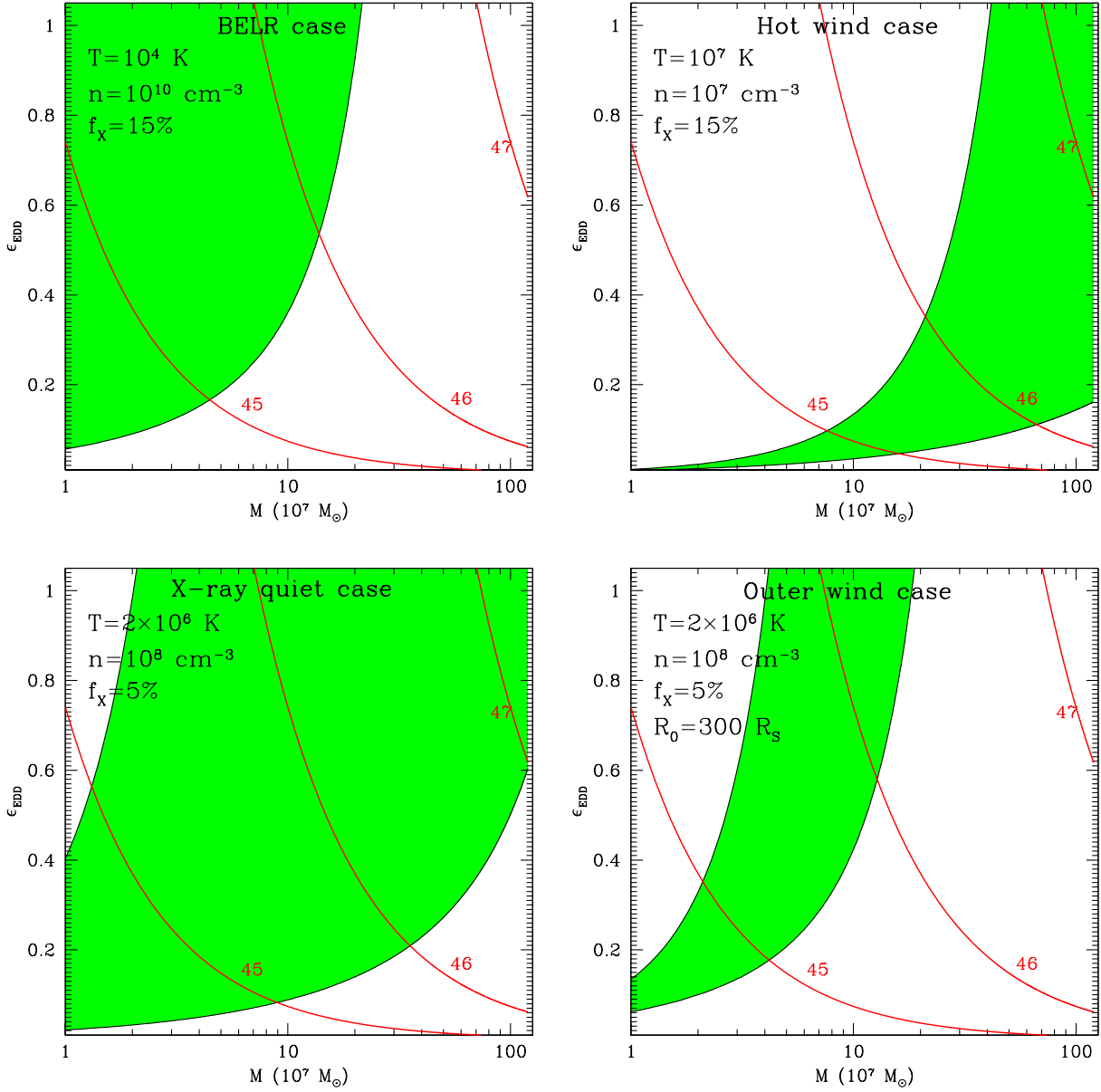} }
\caption{Results of our numerical analysis for several choices of the
initial parameters. The red lines show the total luminosity of the
source. The green (shaded) regions are those allowed 
for wind launching. Upper left panel: BELR case; upper right panel: 
hot wind case; lower left panel: wind from an X-ray quiet source;
lower right panel: wind with a large inner radius ($R_{IN}=300~R_G$).
\label{theory2}}
\end{figure*}

We show in Fig.~\ref{theory2} the allowed regions in four cases,
obtained from the baseline case changing one or two initial parameters,
in order to test four different physical situations: \\
{\bf (a) BELR wind}:
It is interesting to
investigate whether BEL gas could be accelerated directly by the
radiation force. 
%Using $n\sim 10^{10}$~cm$^{-3}$ (constant over R) and $T\sim 10^4$~K, we
%obtain the results plotted in
Fig.~\ref{theory2}a, shows that BEL gas cannot be accelerated to high
velocities by the radiation force, except for relatively low mass black
holes ($M_7 < 10$) and high Eddington ratios (excpet for the lowest masses). 
The physical reason is
that in an overly dense gas the Doppler shift due to the radial
acceleration is insufficient to prevent the gas from being
self-shielded against line absorption.  Thus in a too dense gas the
effective optical depth, $t$, is never small enough to make the force
multiplier $M(t)$ much higher than 1, for reasonable values of the
radial velocity gradient ($\frac{dv_r}{dr}>10^{-6}$~s$^{-1}$).
We stress that this result does not imply that a wind
containing a cold and dense phase is possible only at low masses.
Indeed, if the BEL gas is confined by a hot wind, it is expected that the
pressure of the warm gas would drive them to similar
velocities. However, a more detailed study is needed to test this
statement. Here we only note that in a wind like the one described in
Elvis (2000) most of the kinetic energy is in the warm phase
(80-90\%).  Therefore, from an energetic point of view it is likely
that the dynamics of the whole wind (cold phase, i.e. the BEL gas,
and warm phase) is determined by the warm phase.

{\bf (b) Hot wind:} Fig.~\ref{theory2}b shows the allowed region for
a hot, low density wind. Only at the highest masses can such a wind be launched,
because of the over-ionization due the low wind density.
The ionization parameter of the wind decreases at high BH masses 
because in our scheme the distance of the stream lines is  always the
same in units of $R_S$. Therefore, increasing the BH mass implies
a linear increase of the luminosity (at a given $\epsilon_{EDD}$)
and a quadratic increase of the physical distance, resulting in a 
linear decrease of the ionization parameter with $M_{BH}$. 

We note that this case is the same as in the second {\em QWIND} survey (Fig.~10).
Again, the allowed region obtained with the analytic approach is
larger than that predicted with {\em QWIND} (Fig.~\ref{survey2}).
 
{\bf (c) X-ray weak source:} The fraction of the bolometric luminosity
emitted in the X-rays, $f_X$, is important in the determination of
the allowed region for a wind. The value used in the
standard parameters set, $f_X$=15\%, is in
agreement with the observed Spectral Energy Distributions (SED) of PG
quasars (Elvis et al. 1994, Laor et al. 1997). However $f_X$ depends
on luminosity (Zamorani et al. 1981, Yuan et al. 1998, Steffen et al.~2006) 
and so higher
values, $\sim20-25$\% are found in nearby Seyfert galaxies. High
values are also found in radio loud quasars (in this case however
beaming could play a role).  On the other side, lower values,
$\sim$5\% are found in high luminosity (L$_{2500\AA}>
10^{31}$~erg~s$^{-1}$~Hz$^{-1}$, Yuan et al. 1998) and high redshift
(z$>$4, Vignali et al. 2002, Steffen et al.2006) quasars. 
In Fig.~\ref{theory2}c we show the case for an X-ray quiet source,
with $f_X=5$\% and the other parameters as in the baseline study.
The allowed region is increased, because 
lower X-ray emission implies a lower ionization parameter.
This is important especially at low BH masses, where the requirements
to avoid gas over-ionization are more stringent.

{\bf (d) Wind Inner Radius:} 
We changed the inner wind radius from 
$100 R_S$ to $300 R_S$ (Fig.~\ref{theory2}d). 
The allowed wind region is smaller than in our baseline study, because
part of the parameter space is excluded both at high BH masses
(as is apparent from a comparison with Fig.~\ref{theory1}),
where a wind cannot be launched because the
physical distance of the gas is too high for the radiation pressure 
to be effective. 

Summarizing, 
our results indicate that only a limited range of physical conditions
can give rise to a radiation driven wind in AGNs, but also that these
conditions do exist, and are the same as those observed in AGN warm
absorbers, and needed for a warm gas that pressure confines the BEL gas.

Furthermore, we showed how our simple approach can provide 
an immediate physical understanding of the dependence of the wind on
the individual initial parameters. 

The analytical results presented here have several important limitations, as
noted at the start of Section~4. In addition, our treatment is based on a
condition of existence for the solutions of the equation of motion of
a wind element. This is a less stringent condition than requiring an
actual wind solution for a given set of initial parameters.  As a
result, real wind solutions are possible for smaller parameter regions
than those shown in Figs.~\ref{theory1} and~\ref{theory2}.
Importantly though, we do not expect any real solution outside
these regions. The exception is  for $\epsilon_{EDD}\geq~1$, for which a wind can
be accelerated through electron scattering alone.

%%%%%%%%%%%%%%%%%%%
%%%%%%%%%%%%%%%
%%%%%%%%%%%%%%%

%%%%%%%%%%%%%%%%%%%%%%%%%%%%%%%%%%%%%%%%%%%%%%%%%%%%%%%%%%%%%%%%%%%%%%%%%
%\section{Discussion}

%We chose the SS disc because it is the simplest and best known disc
%structure. With this choice we were able to obtain a wind with the
%characteristics required by several pieces of observational
%evidence. It is important to check whether such solutions are possible
%with significantly different density, of if instead only initial
%parameters close to those used in our examples can give wind
%solutions.  In the latter case, our results would provide an indirect
%support for the optically thick, geometrically thin disc structure
%postulated in the SS model.  

%%%%%%%%%%%%%%%%%%%%%%%%%%%%%%%%%%%%%%%%%%%%%%%
\section{Comparison with previous results}

A comparison of our results with those obtained with more complex
hydrodynamical codes is useful to test the reliability of 
{\em QWIND}.  In particular, we refer to the work of P00.  in which a
radiation-driven wind from a SS disc is studied, solving the wind
equations through a 2-dimensional hydrodynamical code.

The properties of the solutions are remarkably similar.
In particular:

\begin{itemize}
\item The velocity profile in P00 (their Fig. 2) 
%is different in the
%vertical part (we assume a quite large initial vertical velocity) but
is remarkably similar after the wind has bent towards the radial
direction.  In particular, when the external force becomes effective,
and the gas changes its direction, the velocity very rapidly increases
up to its maximum value (around 10$^4$~km~s$^{-1}$). This is in
agreement with our main approximation, i.e. the separation between a
internal pressure dominated phase and an external radiation force
dominated phase.

\item The ionization parameter profiles are quite similar. This is not
surprising, and indeed is only a check of self-consistency, since we
have adopted the same assumptions for the X-ray cross-section as
P00. As discussed above, in this scenario the ionization parameters
rapidly drops from large ($>10^5$) to very small ($<10^3$) values
across a thin layer of $N_H\sim 10^{22}$~cm$^{-2}$.
\item The final values for the ejected mass are comparable.
%Using our Eq.~16, 
We find for our ``case study'' a value of $\sim~1~M_\odot$~yr$^{-1}$
%(this is $\sim10$ times higher than in the case discussed in P00,
%however our mass is 3 times higher, and $\dot{M}_D \propto
%M_\bullet^2$).
close to that of P00 for the same black hole mass (we
note that the mass loss rate scales as $M_{BH}^2$).
% Considering that most of the gas falls back into the
%disc, the actual rate of mass loss (i.e. the mass of the gas which
%reaches the escape velocity) is in our model
%$\sim0.2~M_\odot$~yr$^{-1}$.

%\item 
%The density and temperature profiles are significantly
%different.  The initial density in the model of P00 is not a free
%parameter, and is determined by assuming the vertical structure of a
%SS disc.  However, 
%The initial density values in our model, and the
%density of the fast outflowing gas is comparable with the results of
%P00.  The temperature we assume is instead significantly higher (by a
%factor $\sim100$) than in P00.  This affects the results through the
%dependence of the effective optical depth on the temperature
%(Eq. 1). A factor 100 lower temperature would imply a $\sim 4$ times
%higher force multiplier (in the non-asymptotic regime).  In this
%regime, it would be much easier to launch a wind. In particular, we do
%not expect the inner radius of the solutions to change (since it is
%determined by the ionization balance), while the outer radius should
%become larger, being determined by the balance between gravity and
%radiation force. The final velocities of the wind would also increase.
%However, this scenario does not correspond to any know observed
%feature in quasars spectra: BEL gas  has similar
%temperatures but much higher densities (see discussion in Sect. 3),
%while the ``warm absorber'' has temperatures of the order of a few
%10$^6$~K or lower.  Therefore, we will not investigate this kind of
%solution further.

\end{itemize}

Summarizing, the comparison with the results of P00 shows that our
model is in qualitative agreement with a much more complex and more
precise treatment. Given the huge uncertainties in the initial
conditions (both regarding the structure of the disc and the mechanism
of initial launching of the gas) we believe our approach is well
supported.  We remind the reader however that our simplified approach should be
used as a way to quickly explore and physically understand the
possible solutions, while a more detailed treatment is needed if one
is interested in the precise values of the parameters involved or in more
detailed structure.

%%%%%%%%%%%%%%%%%%%%%%%%%%%%%%%%%%%%%%%%%%%%%%

%SCHEME: 
%(1) The outflow has the same properties of BAL outflows, and
%also the right covering angle 
%
%(2) It is natural to link the BELC with the outflowing wind (right
%distance, right outflow velocity if compared with BAL)
%
% - Case for axisymmetric BELCs: No Lyman edge (Maiolino et al. 2001)\\
%
%Fast variability (our work of 2002, new results on N4151) which means
%cold gas at BELC distance in Type 2s but not in Type 1s\\
%
%Others (there are recent papers claiming this, but I cannot
%remember... do you?) 
%3) A more detailed structure for a wind is the one proposed in Elvis (2000).
%This model is in agreement with the two points above, and in addition it can
%explain NALs and reflection, assuming that the wind start vertically, and
%is thin ($\Delta R/R\sim0.1$, transverse$N_H\sim10^{22}$~cm$^{-2}$).
%The structure of our solution is in overall agreement with this
%model, however, the results in Fig.~\ref{survey3} show that such a
%thin wind is possible within a particular range of initial parameters
%(in the example of our survey, for $\epsilon_{EDD}\sim0.4$.....

%%%%%%%%%%%%%%%%%%%%%%%%%%%%%%%%%%%%%%%%%%%%%%%%%%%%%%%%%%%%%%%%%%%%%%%%%
\section{Conclusions, and Future Work}

We have shown that a radiation-driven wind can be accelerated to
velocities of about 10$^4$~km~s$^{-1}$ from a Shakura-Sunyaev
accretion disc at a distance from the central black hole of the order
of that of the BEL clouds, for densities and temperatures in the range
observed in quasars warm absorbers.

Thanks to our simple approach, we were able to make an analytic
estimate of the parameter space in which a wind solution is allowed,
and to build a fast code, {\em QWIND}, which can be used to better
investigate the dependence on the several unknown initial parameters.

The main limitation of our study is the assumption of a substantial
initial velocity (a few 100~km~s$^{-1}$) which
makes the wind supersonic from the beginning.  We did not discuss the
physical mechanism through which the disc provides the initial kinetic
energy to the gas.
%, but we showed that this is in agreement with what found when
%process does not require a
%particularly high efficiency (i.e. the energy radiated by each ring of
%the disc is usually much greater than the energy transferred to the
%gas, Appendix A).  

Our main results are: 
\begin{enumerate}
\item Our model reproduces the global results of more complex hydrodynamical
codes.  This makes us confident that our approach can be used to
explore the huge space of initial conditions, in order to understand
in which cases a wind can be launched.  The results of our analytical
study are in agreement with those of the numerical code {\em QWIND}, in the sense
that parameters space regions excluded by the analytical analysis are
also excluded with the more precise numerical approach.
% (except for
%very high values of $\epsilon_{EDD}$, see discussion in Section
%4.2). This gives us confidence that the essential physics is captured in our
%simple treatment. 

\item For initial parameters typical of AGN warm absorbers, a wind can
arise only for relatively high accretion efficiencies
($\epsilon_{EDD}>0.3$) and in a {\em narrow range of disc radii}. This
is due to overionization at too small radii, and to a too small UV
radiation and initial velocity at too large radii, and is an inherent
feature of radiation driven disc winds in AGNs.

\item The inner part of the gas, which is too overionized to be 
effectively accelerated, forms a ``failed wind'' which
removes the X-ray radiation from directions close to the disc plane, 
allowing a fast decrease of the 
ionization parameter with the distance form the central source, 
and so allowing the line driven wind acceleration
at larger radii. This component is analogous to the
``hitchhiking gas'' of Murray \& Chiang~(1995).

\item We have shown how QWIND can be used to gain a simple
physical understanding of the dependence of the wind properties on
the initial parameters, such as the X-ray to UV ratio, the Eddington ratio,
the initial density and temperature, the central black hole mass.
Some of the results are at first counter-intuitive, as, for example,
the non-monotonic dependence of the wind final velocity on the
Eddington ratio: in some cases more luminous sources can have slower winds,
with the other physical parameters being held constant.

\item The terminal velocity of the wind is typically of the order of
$1-2~10^4~$km~s$^{-1}$, and the cone covering factor is a few percent of
the lines of sight. This is in reasonable agreement with observations of BAL
quasars.

\item The wind angle above the disc is substantial ($\sim20$deg), and so is
neither equatorial nor polar. Some $\sim35$\% of the viewing directions will see the
X-ray and UV continuum source through the wind.

%\item {\sc More?}

\end{enumerate} 

Our findings could have deep implications for wind-embedded BEL
models, like that of Elvis (2000).  Black holes accreting in
conditions outside the narrow allowed ranges would not have winds, and
so neither absorption lines, nor the BELs would be
observed. Such 'naked quasars' would be picked up from their
non-stellar colors in the Sloan Digital Sky Survey or by their high
X-ray flux in X-ray surveys. A few ``lineless'' quasars have been found but 
they are rare (e.g. McDowell et al.~1995). If AGN winds are radiation-driven, then
some self-regulation mechanism must be involved in the accretion process,
to keep the AGNs at the conditions needed for a radiation driven wind.

Our main aims at this stage were to check the reliability of our
approach, and to give significant examples of its use. The next major
step, which will be presented in a forthcoming paper, is a systematic
survey of initial parameters, to give a more general view of the
physical conditions needed to accelerate winds by line driving. In
particular, we will investigate how the results are affected by
changes in the wind inner radius and in the initial density and
velocity profiles. 

\section*{Acknowledgments}
%We thank Colin Norman for a careful reading of the manuscript. 
We are grateful to Jonathan Mc Dowell for his help with the numerical
code, and to the anonymous referee for his/her comments and
careful check of the text, which greatly improved the clarity of the work. 
%We thank the anonymous referee for constructive criticism which
%significantly improved the paper.  
%
This work has been partially supported by NASA grants NNX08AX78G, and NNX07AI22G,
and by grant ASII-INAF I/088/06/0.
%\email{aastex-help@aas.org}.

%%%%%%%%%%%%%%%%%%%%%%%%%%%%%%%%%%%%%%%%%%%%%%%%%%%%%%%%%%%%%%%%%%%%%%%%%

%%%%%%%%%%%%%%%%
\end{document}